\documentclass[journal,12pt, draftclsnofoot, onecolumn]{IEEEtran}

\usepackage{graphicx}
\usepackage{amsmath,amssymb,dsfont,bbm,epstopdf,pgfplots,algorithm, algpseudocode , graphicx} 
\usepackage{color}
\usepackage{cite}
\usepackage{graphics}
\usepackage{tikz}
\usepackage{pgfplots}
\usepackage{caption}
\usepackage{subcaption}
\usetikzlibrary{shapes, arrows, decorations.markings, arrows.meta}
\usetikzlibrary{patterns} 
\allowdisplaybreaks
\graphicspath{{.}{./Figures/}} 
\allowdisplaybreaks[4] 

\newtheorem{lemma}{Lemma}
\newtheorem{corollary}{Corollary}
\newtheorem{theorem}{Theorem}
\newtheorem{proposition}{Proposition}
\newtheorem{remark}{Remark}

\newcommand{\D}{\text{D}}
\newcommand{\Dt}{\D_{\text {Tx}}}
\newcommand{\Dr}{\D_{\text {Rx}}}
\renewcommand{\Pr}{{\mathbb{P}}}

\newcommand{\sRkf}{{\{1,\ldots,\lfloor 2^{n R_k^{(F)}} \rfloor \}}}
\newcommand{\sRks}{{\{1,\ldots,\lfloor 2^{n R_k^{(S)}} \rfloor \}}}

\newcommand{\muT}{{\mu}_\Tx}
\newcommand{\muR}{{\mu}_\Rx}

\renewcommand{\S}{{\sf{S}}}

\newcommand{\Ntk}{\mathcal{N}_{\textnormal{Tx}}(k)}
\newcommand{\Nrk}{\mathcal{N}_{\textnormal{Rx}}(\tk)}

\renewcommand{\i}{{\iota}}

\newcommand{\K}{[K]}

\newcommand{\vect}[1]{\boldsymbol{#1}}

\newcommand{\Tx}{\textnormal{Tx}}
\newcommand{\Rx}{\textnormal{Rx}}

\newcommand{\MkF}{M_k^{(F)}}
\newcommand{\MkS}{M_k^{(S)}}

\definecolor{ForestGreen}{rgb}{0.0, 0.5, 0.0}
\newcommand{\mw}[1]{{\color{black}#1}}
\newcommand{\maw}[1]{{\color{black}#1}}

\newcommand{\fillaround}[3][fill=yellow]{
\node[filledhex,#1] at (#3,#3+4) {};
}
\renewcommand{\P}{\mathsf{P}}
\renewcommand{\L}{\mathsf{L}}

\newcommand{\Ds}{ \left \lfloor \frac{\D}{2}  \right \rfloor}

\newcommand{\Drss}{ \left \lfloor \frac{\Dr-1}{2}  \right \rfloor}

\newcommand{\Dtss}{ \left \lfloor \frac{\Dt-1}{2}  \right \rfloor}

\newcommand{\mtbone}{\mu_{\Tx, \text{both}}^{(\text{r})}}
\newcommand{\mrbone}{\mu_{\Rx, \text{both}}^{(\text{r})}}
\newcommand{\mtbtwo}{\mu_{\Tx, \text{both}}^{(\text{t})}}
\newcommand{\mrbtwo}{\mu_{\Rx, \text{both}}^{(\text{t})}}

\newcommand{\mtsone}{\mu_{\Tx, \S}^{(\text{t})}}

\newcommand{\mrstwo}{\mu_{\Rx, \S}^{(\text{r})}}

\newcommand{\mtbonew}{\mu_{\Tx, \text{both}}^{(\text{r})}}
\newcommand{\mrbonew}{\mu_{\Rx, \text{both}}^{(\text{r})}}
\newcommand{\mtbtwow}{\mu_{\Tx, \text{both}}^{(\text{t})}}
\newcommand{\mrbtwow}{\mu_{\Rx, \text{both}}^{(\text{t})}}

\newcommand{\mtsonew}{\mu_{\Tx, \S}^{(\text{t})}}

\newcommand{\mrstwow}{\mu_{\Rx, \S}^{(\text{r})}}

\newcommand{\mtboneh}{\mu_{\Tx, \text{both}}^{(\text{r})}}
\newcommand{\mrboneh}{\mu_{\Rx, \text{both}}^{(\text{r})}}
\newcommand{\mtbtwoh}{\mu_{\Tx, \text{both}}^{(\text{t})}}
\newcommand{\mrbtwoh}{\mu_{\Rx, \text{both}}^{(\text{t})}}

\newcommand{\mtsoneh}{\mu_{\Tx, \S}^{(\text{t})}}

\newcommand{\mtstwoh}{\mu_{\Tx, \S}^{(\text{r})}}
\newcommand{\mrstwoh}{\mu_{\Rx, \S}^{(\text{r})}}

\newcommand{\mtbones}{\mu_{\Tx, \text{both}}^{(\text{r})}}
\newcommand{\mrbones}{\mu_{\Rx, \text{both}}^{(\text{r})}}

\newcommand{\mrstwos}{\mu_{\Rx, \S}^{(\text{r})}}

\newcommand{\tk}{k}

\newcommand{\q}{\mathcal Q}

\newcommand{\Tf}{{\mathcal T_{\text{fast}}} }
\newcommand{\Ts}{{\mathcal T_{\text{slow}}} }

\newcommand{\sfbw}{\S^{(F)}_{\text{both}}}
\newcommand{\ssbw}{\S^{(S)}_{\text{both}}}
\newcommand{\ssmw}{\S^{(S)}_{\text{max}}}
\newcommand{\sfmw}{\S_{\text{no-coop}}}
\newcommand{\ssncw}{\S_{\text{no-coop}}}

\newcommand{\sfbh}{\S^{(F)}_{\text{both}}}
\newcommand{\ssbh}{\S^{(S)}_{\text{both}}}
\newcommand{\sfmh}{\S_{\text{no-coop}}}
\newcommand{\ssmh}{\S^{(S)}_{\text{max}}}
\newcommand{\ssnch}{\S_{\text{no-coop}}}

\newcommand{\sfbsr}{\S^{(F)}_{\text{both}}}
\newcommand{\ssbsr}{\S^{(S)}_{\text{both}}}

\newcommand{\ssmsr}{\S^{(S)}_{\text{max}}}

\newcommand{\ssms}{\S^{(S)}_{\text{max}}}

\begin{document}

\title{Coordinated Multi Point Transmission and Reception for Mixed-Delay Traffic}

\author{\IEEEauthorblockN{Homa Nikbakht, ~\IEEEmembership{Student  Member, IEEE},  Mich\`ele Wigger, ~\IEEEmembership{Senior Member, IEEE}, and Shlomo Shamai (Shitz), \IEEEmembership{Fellow, IEEE }}

\thanks{H. Nikbakht and M.   Wigger are with  LTCI, T$\acute{\mbox{e}}$l$\acute{\mbox{e}}$com Paris, IP Paris, 91120 Palaiseau, France. E-mails: \{homa.nikbakht, michele.wigger\}@telecom-paris.fr. S. Shamai, is with the Department of Electrical Engineering, Technion---Israel Institute of Technology, Technion City, 32000, Israel, e-mail: sshlomo@ee.technion.ac.il. Part of this work has been presented at IEEE SPAWC 2019  \cite{HomaSPAWC2019}.
The work of H. Nikbakht and M. Wigger has been supported by the European Union's Horizon 2020 Research And Innovation Programme, grant agreement no. 715111. The work of S. Shamai has been supported by the European Union's Horizon 2020 Research And Innovation Programme, grant agreement no. 694630.}
}
%

\maketitle

\IEEEpeerreviewmaketitle


\begin{abstract}
This paper analyzes the multiplexing gains (MG) for simultaneous transmission of delay-sensitive and delay-tolerant data 
over interference networks. In the considered model, only delay-tolerant data can profit from coordinated
multipoint (CoMP) transmission or reception techniques, because delay-sensitive data has to be
transmitted without further delay. Transmission of delay-tolerant data is also subject to a delay
constraint, which is however less stringent than the one on delay-sensitive data. Different coding schemes
are proposed, and the corresponding MG pairs for delay-sensitive and delay-tolerant data
are characterized for Wyner's linear symmetric network and for Wyner’s two-dimensional hexagonal
network with and without sectorization. For Wyner's linear symmetric  also an information-theoretic
converse is established and shown to be exact whenever the cooperation rates are sufficiently large or the delay-sensitive MG is
small or moderate. These results show that on Wyner's symmetric linear network and for sufficiently
large cooperation rates, the largest MG for delay-sensitive data can be achieved
without penalizing the maximum sum-MG of both delay-sensitive and delay-tolerant data. A similar
conclusion holds for Wyner’s hexagonal network only for the model with sectorization. In the model without sectorization, a penalty in sum-MG is incurred whenever one insists on a positive delay-sensitive MG.

\end{abstract}

\section{Introduction}

One of the main challenges for  future wireless communication systems is to accommodate  heterogeneous data streams with different delay constraints. This is also  the focus of various recent works, notably \cite{Kassab2018, Matera2018, Anand2020,  HomaITW2019, shlomo2012ISIT,Zhang2005,Zhang2008}.
In particular, \cite{Kassab2018,Matera2018} study a cloud radio access network (C-RAN) under mixed-delay-constraints traffic. Specifically,   users close to  base stations (BS) transmit delay-sensitive data, which  is directly decoded at the BSs, and users that are further away send delay-tolerant data, which  is decoded at the central processor.
 In this paper we refer to delay-tolerant data as  \emph{``slow" messages}, and to delay-sensitive data as \emph{``fast" messages}. In \cite{HomaITW2019}, we extended above C-RAN model to allow each user to send  both  ``fast"  and ``slow" messages, \mw{and to time-varying fading channels. 
The  results in \cite{HomaITW2019} show 
that in most regimes,    the stringent delay constraint on ``fast" messages  penalizes the overall performance (sum-rate) of the system. 

The work in  \cite{shlomo2012ISIT} proposes a \mw{superposition} approach over a fading channel to   communicate  ``fast" messages within single coherence blocks and  ``slow" messages  over   multiple blocks. In \cite{Zhang2005} a scheduling algorithm is proposed for a $K$-user broadcast network that gives  preference  to the communication of ``fast" messages over  ``slow" messages. A related work was performed in  \cite{Zhang2008}, where  ``fast'' messages can be stored  in a buffer during a single scheduling period.}

The focus of the current work is on the benefits of cooperation for mixed-delay traffics, assuming that only the transmissions of ``slow'' messages  can profit from cooperation between terminals, but not  ``fast'' messages. 
Networks with transmitter- (Tx) and/or receiver- (Rx) cooperation have been considered in many recent works including \cite{Steinberg, Aktas2008, Bross2008, Sendonaris2003, Sendonaris20032, ElGamal2014, ElGamal2014J, Bande19, Shamai2011, Lapidoth2014, Wiggeretal} but mostly  only with a single type of messages, namely the messages that we call ``slow'' messages.   Huleihel and Steinberg  \cite{Steinberg} considered two types of messages: one type that has to be decoded  whether or not the Rx-cooperation link is present, and the other that only has to be decoded when the cooperation  link is present. Inspired by  this model, we studied Wyner's soft-handoff model \cite{Wyner-94,Hanly-Whiting-93} with mixed-delay traffics  in \cite{HomaEntropy2020}, where  the Tx-cooperation messages can only depend on the ``slow" messages in the system and not on the ``fast" messages, and ``fast" messages have to be decoded prior to the Rx-cooperation phase, whereas ``slow" messages can be decoded thereafter. Moreover,  in  \cite{HomaEntropy2020} the total number of Tx- and Rx- cooperation rounds is constrained also for the ``slow" messages as proposed \cite{Wiggeretal}. The results in \cite{HomaEntropy2020} show that, in the high signal to noise ratio (SNR) regime, when both the Txs and the Rxs can cooperate, and for sufficiently large cooperation rates, it is possible to accommodate the largest possible rate for ``fast''  messages without penalizing the maximum sum-rate of both ``fast'' and ``slow'' messages.  When only Txs or only Rxs can cooperate, transmitting also ``fast'' messages causes no penalty on the sum-rate at low ``fast'' rates, but the
sum-rate decreases linearly at high ``fast'' rates.
Notice that the standard approach to combine the transmissions of ``slow'' and ``fast'' messages is to time-share (schedule) the transmission of ``slow'' messages with the transmission of ``fast'' messages. In this approach, the sum-rate decreases   linearly with the rate of the ``fast'' messages and attains  the maximum sum-rate   only  when no ``fast'' messages are transmitted.

The focus of this paper is on the pairs of \emph{Multiplexing Gains (MG)}, also called degrees of freedom or capacity prelogs, that are simultaneously achievable for  ``fast'' and ``slow'' messages.  Using interference alignment with infinite symbol-extensions, the MG of any non-cooperative interference network with $\L$-antenna receivers and sufficiently many transmit antennas is $\L/2$ \cite{Jafar, Sridharan2015}. Such interference alignment techniques however cannot be implemented in practice \cite{Caire2015}, and therefore here we focus on more practical  successive interference cancellation and precoding techniques with Txs and Rxs that can cooperate during a  limited number of interaction rounds. 

We propose a general coding scheme for any interference network with Tx- and Rx-cooperation that simultaneously accommodates the transmissions of ``slow'' and ``fast'' messages,  
and characterize their achievable MG pairs for two  specific cellular network models: Wyner's linear symmetric model \cite{Wyner-94,Hanly-Whiting-93} and Wyner's two-dimensional hexagonal model \cite{Wyner-94} \mw{with and without sectorization.}  For Wyner's symmetric network we also provide an information-theoretic converse result. It matches the proposed set of achievable MG pairs when the cooperation links are of sufficiently high prelogs or when the MG of ``fast" messages is small. These results show that  when the prelog of the cooperation links is sufficiently large, for Wyner's linear symmetric model, \mw{as  for Wyner's linear soft-handoff model \cite{HomaEntropy2020},   it is possible to accommodate the largest possible MG for ``fast''  messages   without penalizing the maximum sum MG of both ``fast'' and ``slow'' messages. Our achievable schemes suggest that the same  also holds for the sectorized hexagonal model considered in this paper where each cell is divided into three non-interfering sectors by employing directional antennas at the BSs \cite{HomaSPAWC2019}. In contrast, for the considered non-sectorized hexagonal model, there seems to be a penalty in maximum sum MG whenever the  ``fast" MG is larger than 0.}

To achieve the described  performances, in our coding scheme, we identify a set of the Txs whose signals  do not interfere. The chosen Txs  send ``fast'' messages and the others send ``slow'' messages or nothing. Communication of  ``fast'' messages is thus only interfered by transmissions of ``slow'' messages and this interference can be described during the Tx-conferencing phase and precanceled at the ``fast'' Txs. Also, ``fast'' Rxs decode their messages immediately and can describe their decoded messages during the Rx-conferencing phase to their adjacent ``slow'' Rxs allowing them to subtract the interference from ``fast'' messages before decoding their own ``slow'' messages. As a result, ``fast'' messages can be decoded based on interference-free outputs and moreover, they do not disturb the transmission of ``slow''  messages. CoMP  transmission or reception   \cite{ElGamal-Book, Annapureddy2015} for limited clusters is then employed to convey the ``slow'' messages.

\subsection{Organization}
The rest of this paper is organized as follows. We end this section with some remarks on notation. The following Sections \ref{setup} and \ref{coding} consider general interference networks and  describe the problem setup and the proposed coding scheme and its multiplexing gain region for such a general network. Sections \ref{sec:sym}--\ref{sec:sec}  specialize the results to the symmetric linear Wyner model and to the  two-dimensional hexagonal
 Wyner model.  Section~\ref{sec:conclusion} concludes the paper. 

\subsection{Notation}
We use the shorthand notations ``Rx" for  ``Receiver" and ``Tx" for ``Transmitter". The set of all integers is denoted by  $\mathbb Z$, the set of positive integers by $\mathbb Z ^{+}$ and the set of real numbers by $\mathbb R$. For other sets we use calligraphic letters, e.g., $\mathcal{X}$.  Random variables are denoted by uppercase letters, e.g., $X$, and their realizations by lowercase letters, e.g., $x$. For vectors we use boldface notation, i.e., upper case boldface letters such as $\mathbf{X}$  for random vectors and lower case boldface letters such as $\mathbf{x}$ for deterministic vectors.) Matrices are depicted with sans serif font, e.g., $\mathsf{H}$. We use   $[K]$ to denote the set $\{1, \ldots, K\}$. We also write $X^n$ for the tuple of random variables $(X_1, \ldots, X_n)$ and  ${\mathbf X}^n$ for the tuple of random vectors   $( \mathbf{X}_1, \ldots,\mathbf{X}_n)$.

\section {Problem Description} \label{setup}

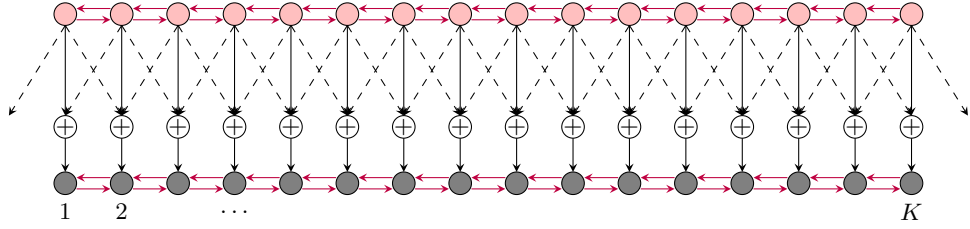
\begin{figure}[t]
  \centering
  \small
\begin{tikzpicture}[scale=1.5, >=stealth]
\centering
\tikzstyle{every node}=[draw,shape=circle, node distance=0.5cm];
\foreach \j in {0,1,...,15} {
 \draw [fill= pink](-3.5 + 0.5*\j, 2) circle (0.1);
  \draw [fill= gray](-3.5 + 0.5*\j, 0.5) circle (0.1);
\node[draw =none] (s2) at (-3.5+ 0.5*\j,1 ) {\footnotesize$+$};
\draw (-3.5 +0.5*\j, 1) circle (0.1);
\draw   [->] (-3.5+ 0.5*\j,1.9)-- (-3.5+ 0.5*\j,1.1);
 \draw   [->] (-3.5+ 0.5*\j,0.9)-- (-3.5+ 0.5*\j,0.1+0.5);
  \draw   [->, dashed] (-3.5+ 0.5*\j,1.9)-- (-3.5+ 0.5*\j + 0.5,1.1);
   \draw   [->, dashed] (-3.5+ 0.5*\j,1.9)-- (-3.5+ 0.5*\j - 0.5,1.1);
}
\foreach \j in {0,1,...,14} {
\draw[->, purple] (-3.5 +0.5*\j+0.1, -0.05+0.5) --(-3.5 +0.5*\j + 0.5-0.1, -0.05+0.5);
\draw[<-, purple] (-3.5 +0.5*\j+0.1, +0.05+0.5) --(-3.5 +0.5*\j + 0.5-0.1, +0.05+0.5);

\draw[->, purple] (-3.5 +0.5*\j+0.1, -0.05+2) --(-3.5 +0.5*\j + 0.5-0.1, -0.05+2);
\draw[<-, purple] (-3.5 +0.5*\j+0.1, +0.05+2) --(-3.5 +0.5*\j + 0.5-0.1, +0.05+2);
 }
 \node[draw =none] (s2) at (-3.5+ 0.5*0,-0.25+0.5 ) {\footnotesize$1$};
 \node[draw =none] (s2) at (-3.5+ 0.5*1,-0.25+0.5 ) {\footnotesize$2$};
 \node[draw =none] (s2) at (-3.5+ 0.5*3,-0.25+0.5 ) {\footnotesize$\ldots$};
 \node[draw =none] (s2) at (-3.5+ 0.5*15,-0.25+0.5 ) {\footnotesize$K$};

\end{tikzpicture}
\vspace*{-2ex}
 \caption{Wyner's symmetric  network.  Black  dashed arrows show interference links and  purple arrows cooperation links. }
  \label{fig3-t2}
 \vspace*{-4ex}
\end{figure}

Consider a cellular  interference network with $K$ cells each consisting of one Tx and Rx pair.  Txs and  Rxs are equipped with $\L$ antennas and  we assume a regular interference pattern except at the network borders. As an example, Fig.~\ref{fig3-t2} shows Wyner's symmetric network where each  cell corresponds to a Tx/Rx pair and the interference pattern is depicted with black dashed lines. 

Each Tx~$k \in [K]$ sends a pair of independent  messages  $M_k^{(F)}$ and $M_k^{(S)}$ to  Rx~$k \in [K]$. The ``fast"  message $\MkF$ is uniformly distributed over the set $
\mathcal{M}_{k}^{(F)} \triangleq \sRkf$ and needs to be decoded subject to a  stringent delay constraint, as we explain shortly.  The ``slow"  message $\MkS$ is uniformly distributed over $\mathcal{M}_{k}^{(S)} \triangleq \sRks$ and is subject to a less stringent decoding delay constraint.  Here, $n$ denotes the blocklength of transmission and $R_k^{(F)}$ and $R_k^{(S)}$  the rates of transmissions of the ``fast" and ``slow" messages.

 We consider a cooperation scenario where neighbouring Txs  cooperate during $\Dt>0$ rounds and neighbouring Rxs   during $\Dr>0$ rounds.
The total cooperation delay is constrained: 
\begin{equation}\label{eq:delay}
\Dt +\Dr \leq \D,
\end{equation} 
where $\D \ge 0$ is a given parameter of the system and the values of $\Dt$ and $\Dr$ are design parameters and can be chosen arbitrary such that \eqref{eq:delay} is satisfied. 

To  describe the encoding at the Txs, denote by   $\Ntk$ the  set of all Txs  that have a direct cooperation link with a given Tx~$k\in[K]$. We refer to $\Ntk$ as the \emph{Tx-neighbouring set}  of Tx~$k$.  Neighbouring Txs can communicate to each other during $\Dt>0$ rounds, where this communication can only depend on  ``slow" messages but not on ``fast" messages. In each conferencing round $j\in\{1,\ldots, \Dt\}$,  Tx~$k$ sends  a cooperation message
	$T_{k\to \ell}^{(j)} \Big(M_{k}^{(S)}, \big\{T_{\ell' \to k}^{(1)}, \ldots, T_{\ell'\to k}^{(j-1)} \big\}_{\ell' \in \Ntk}\Big)$
	to Tx~$\ell$ if  $\ell\in \Ntk$.
	 The cooperation communication is assumed  noise-free but  rate-limited:
	\begin{equation}\label{eq:conference_capatx}
	\sum_{j=1}^{\Dt}        H( T^{(j)}_{k\rightarrow  \ell})         \leq  \mu_{\Tx} \cdot \frac{n}{2} \log (\P), \qquad k\in\K,\; \ell \in \Ntk,
	\end{equation}
	for a given  \emph{Tx-conferencing prelog} $\mu_{\Tx}>0$ and  where $H(\cdot)$ denotes the entropy function. 

 Tx~$k$  computes its channel inputs $\mathbf X_k^n = (\vect X_{k,1}, \ldots, \vect X_{k,n}) \in \mathbb R ^{\L \times n}$  as a function of  its ``fast" and ``slow" messages and of  the $\Dt |\Ntk|$ obtained cooperation messages:
\begin{equation}\label{xkn}
\mathbf X_k^n  =  {f}_k^{(n)} \Big( M_{k}^{(F)}, M_{k}^{(S)}, \{T_{\ell' \to k}^{(1)}, \ldots, T_{\ell'\to k}^{(\Dt)} \}_{\ell' \in \Ntk} \Big).
\end{equation}

The channel inputs have to satisfy
the average block-power constraint almost surely:
\begin{equation}\label{eq:power}
\frac{1}{n} \sum_{t=1}^n ||\mathbf X_{k,t}||^2 
\leq \P, 
\quad \forall\ k \in \K.
\end{equation}

To  describe the decoding,  denote the \emph{Rx-neighbouring set}  of a given Rx~$k\in[K]$, i.e.,  the set of all receivers that can directly exchange cooperation messages with Rx~$k$, by $\Nrk$. Also, define the \emph{interference set}  $\mathcal I_{\tk}$ as the the set of all Txs whose signals interfere at  Rx $\tk$.

 Decoding takes place in two phases. During the  \emph{fast-decoding phase}, each  Rx~$\tk$  decodes its   ``fast"  message $M_{\tk}^{(F)} $ based on its  channel outputs $\mathbf Y_{\tk}^n = ( \vect Y_{\tk,1}, \ldots, \vect Y_{\tk,n}) \in \mathbb R^{\L \times n}$, where 
\begin{equation}\label{ykn}
\mathbf Y_{\tk}^n =  \mathsf H_{k,k} \vect X_k^n + \sum_{\hat k \in \mathcal I_{\tk}} \mathsf{H}_{\hat k ,k}  \vect X_{\hat k}^n + \maw{\vect Z_{k}^n,}
\end{equation}
and $\vect{Z}_{k,k}^n$ is  i.i.d. standard Gaussian noise, and  the fixed $\L$-by-$\L$  full-rank matrix $\mathsf{H}_{\hat k,k}$ models the channel from Tx~$\hat k$ to the receiving antennas at Rx $ k$. So, Rx~$\tk$ produces:
\begin{equation}\label{mhatf}
\hat{{{M}}} _{\tk}^{(F)} ={g_{\tk}^{(n)}}\big( \mathbf Y_{\tk}^{n}\big),
\end{equation} 
 using some decoding function $g_{\tk}^{(n)}$ on appropriate domains.
In the subsequent \emph{slow-decoding} \emph{phase},  
 each Rx~$\tk\in \K$  sends  a conferencing message  
$Q^{(j)}_{\tk\rightarrow \ell}\Big( \mathbf Y_k^n,\big\{ Q^{(1)}_{\ell' \rightarrow \tk},\ldots , Q^{(j-1)}_{\ell' \rightarrow \tk}\}_{\ell' \in \Nrk}\big\}\Big)$ during cooperation round  $j\in \{1, \ldots, \Dr \}$ 
to Rx~$\ell$ if  $\ell\in \Nrk$. 
The cooperative communication is noise-free, but rate-limited:
\begin{equation}\label{eq:conference_caparx}
\sum_{j=1}^{\Dr}	H( Q^{(j)}_{\tk\rightarrow \ell })	 \leq   \mu_{\Rx} \cdot \frac{n}{2} \log (\P), \qquad k \in \K, \; \ell \in \Nrk,
\end{equation}
for  given \emph{Rx-conferencing prelog} $\mu_{\Rx}>0$.
 Each Rx~$\tk$ decodes its desired ``slow" message as
\begin{equation}\label{mhats}
\hat{{M}}_{\tk}^{(S)}={b_{\tk}^{(n)}}\Big( \mathbf Y_{\tk}^n, \Big\{ Q^{(1)}_{\ell'\rightarrow \tk}, \ldots, Q^{(\Dr)}_{\ell'\rightarrow \tk}\Big\}_{\ell' \in \Nrk}\Big),
\end{equation}
 using some decoding function  $b_{\tk}^{(n)}$ on appropriate domains. 

 Throughout this article we  assume short range interference and thus:
 \begin{equation}
\mathcal I_{\tk} \subseteq (\Nrk \cap \Ntk ).
\end{equation}

Given power $\P>0$, maximum delay $\D \ge 0$, and  cooperation prelogs  $\mu_{\Rx},\mu_{\Tx}\geq 0$, average rates $(\bar{R}_{K}^{(S)}(\P), \bar{R}_{K}^{(F)}(\P))$ are called achievable, if there exist rates $\{(R_k^{(F)}, R_{k}^{(S)})\}_{k=1}^{K}$ satisfying
\begin{IEEEeqnarray}{rCl}\label{eq:average_rates}
\bar R_K^{(F)} := \frac{1}{K} \sum_{k=1}^K R_{k}^{(F)}, \quad \text{and} \quad 
\bar R_K^{(S)} := \frac{1}{K} \sum_{k=1}^K R_{k}^{(S)} ,
\end{IEEEeqnarray} 
and 
encoding, cooperation, and decoding functions  for these rates satisfying constraints \eqref{eq:delay},   \eqref{eq:conference_capatx}, \eqref{eq:power},   and \eqref{eq:conference_caparx} and so that the probability of error vanishes:
\begin{equation}
p(\textnormal{error}) \triangleq \Pr \bigg[\bigcup_{k \in \K} \Big(\big( \hat{M}_k^{(F)} \neq M_k^{(F)}\big)  \cup \big(\hat{M}_k^{(S)} \neq M_k^{(S)}\big)\Big)  \bigg] \to 0 \quad \textnormal{as } \quad n\to \infty.
\end{equation}

An MG pair  $(\S^{(F)},\S^{(S)})$ is called \emph{achievable}, if for every positive integer $K$  and power $P>0$ there exist achievable average rates   $\{\bar{R}_{K}^{(F)}(\P),\bar R_K^{(S)}(\P) \}_{\P>0}$
satisfying
\begin{IEEEeqnarray}{rCl} \label{sf}
\S^{(F)} \triangleq \varlimsup_{K\rightarrow \infty}\; \varlimsup_{\P\rightarrow\infty} \;\frac{ \bar R_K^{(F)}(\P)}{\frac{1}{2}\log (\P)}, \quad \text{and} \quad 
\S^{(S)} \triangleq \varlimsup_{K\rightarrow \infty}\; \varlimsup_{\P\rightarrow\infty} \;\frac{  \bar R_K^{(S)}(\P)}{\frac{1}{2}\log (\P)}.
\end{IEEEeqnarray}
The  closure of the set of all achievable   MG pairs $(\S^{(F)}, \S^{(S)})$ is called \emph{optimal MG region} and  denoted $\mathcal{S}^\star(\mu_{\Tx},\mu_{\Rx},\D)$.

\section{Coding Schemes and Achievable Multiplexing Gains} \label{coding}
We describe various coding schemes that either transmit both ``fast'' and ``slow'' messages (Subsections~\ref{sub:both} and \ref{sub:both2}) or  only ``slow" messages (Subsection~\ref{sub:onlyslow}), and a scheme that does  not use any kind of cooperation (Subsection~\ref{sub:onlyfast}).  

An important building block in our coding schemes  is CoMP transmission or CoMP reception. Depending on which of the two is used, the scheme requires more Tx- or Rx-cooperation rates. So, depending on the application, any of the two can be advantageous. In some applications, cooperation rates might however be too low to employ either of the two. In this case, the proposed schemes can be time-shared with alternative schemes that require less or no cooperation rates at all. Alternatively, the proposed schemes can be employed with a smaller number of cooperation rounds $\D' <\D$, which also reduces the required cooperation prelog in all our schemes.   
  
\subsection{Coding scheme to transmit both ``fast'' and ``slow'' messages with CoMP reception:} \label{sub:both}
Split the total number of conferencing rounds between Tx- and Rx-conferencing as:
\begin{equation}
\Dt=1 \quad \textnormal{and} \quad \Dr= \D-1.
\end{equation} 

\subsubsection{Creation of subnets and message assignment}
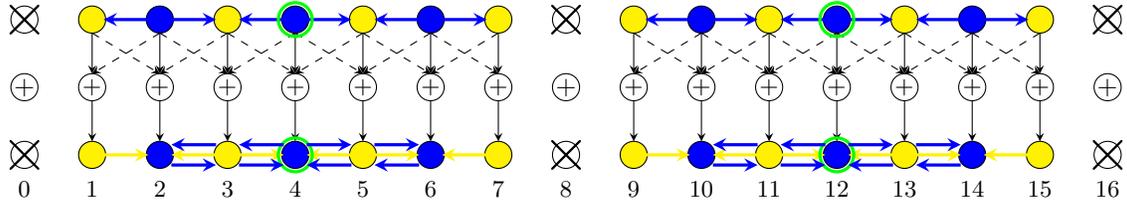
\begin{figure}[t]
  \centering
  \small
    \centering
\begin{tikzpicture}[scale=0.9, >=stealth]
\centering
\tikzstyle{every node}=[draw,shape=circle, node distance=0.5cm];
 \foreach \j in {-1} {
 \draw [fill= white](-3.5 + 1*\j, 2) circle (0.2);
  \draw [fill= white](-3.5 +1*\j, 0) circle (0.2);
\node[draw =none] (s2) at (-3.5+1*\j,1 ) { $+$};
\draw (-3.5 +1*\j, 1) circle (0.2);
\node[draw =none, rotate = 45] (s2) at (-3.5+1*\j,2 ) { \Huge$+$};
\node[draw =none, rotate = 45] (s2) at (-3.5+1*\j,0 ) { \Huge$+$};
 }
 
\foreach \j in {7} {
 \draw [fill= white](-3.5 + 1*\j, 2) circle (0.2);
  \draw [fill= white](-3.5 +1*\j, 0) circle (0.2);
\node[draw =none] (s2) at (-3.5+1*\j,1 ) { $+$};
\draw (-3.5 +1*\j, 1) circle (0.2);
\node[draw =none, rotate = 45] (s2) at (-3.5+1*\j,2 ) { \Huge$+$};
\node[draw =none, rotate = 45] (s2) at (-3.5+1*\j,0 ) { \Huge$+$};
 }
 
 \foreach \j in {1,3,5,9,11,13} {
 \draw [fill= blue](-3.5 + 1*\j, 2) circle (0.2);
  \draw [fill= blue](-3.5 +1*\j, 0) circle (0.2);
\node[draw =none] (s2) at (-3.5+1*\j,1 ) {\footnotesize$+$};
\draw (-3.5 +1*\j, 1) circle (0.2);
 \draw   [->] (-3.5+1*\j,0.8)-- (-3.5+ 1*\j,0.2);
  \draw   [->] (-3.5+1*\j,1.8)-- (-3.5+ 1*\j,1.2);
  \draw   [->, dashed] (-3.5+ 1*\j,1.8)-- (-3.5+ 1*\j + 1,1.2);
   \draw   [->, dashed] (-3.5+ 1*\j,1.8)-- (-3.5+ 1*\j - 1,1.2);
  
 }
 \foreach \j in {0} {
 \draw [fill= yellow](-3.5 + 1*\j, 2) circle (0.2);
  \draw [fill= yellow](-3.5 +1*\j, 0) circle (0.2);
\node[draw =none] (s2) at (-3.5+1*\j,1 ) {\footnotesize$+$};
\draw (-3.5 +1*\j, 1) circle (0.2);
 \draw   [->] (-3.5+1*\j,0.8)-- (-3.5+ 1*\j,0.2);
  \draw   [->] (-3.5+1*\j,1.8)-- (-3.5+ 1*\j,1.2);
  \draw   [->, dashed] (-3.5+ 1*\j,1.8)-- (-3.5+ 1*\j + 1,1.2);
 }
 
 
  \foreach \j in {15} {
 \draw [fill= white](-3.5 + 1*\j, 2) circle (0.2);
  \draw [fill= white](-3.5 +1*\j, 0) circle (0.2);
\node[draw =none] (s2) at (-3.5+1*\j,1 ) { $+$};
\draw (-3.5 +1*\j, 1) circle (0.2);
\node[draw =none, rotate = 45] (s2) at (-3.5+1*\j,2 ) { \Huge$+$};
\node[draw =none, rotate = 45] (s2) at (-3.5+1*\j,0 ) { \Huge$+$};
 }
 
 \foreach \j in {2,4,10,12} {
 \draw [fill= yellow](-3.5 + 1*\j, 2) circle (0.2);
  \draw [fill= yellow](-3.5 +1*\j, 0) circle (0.2);
\node[draw =none] (s2) at (-3.5+1*\j,1 ) {\footnotesize$+$};
\draw (-3.5 +1*\j, 1) circle (0.2);
 \draw   [->] (-3.5+1*\j,0.8)-- (-3.5+ 1*\j,0.2);
  \draw   [->] (-3.5+1*\j,1.8)-- (-3.5+ 1*\j,1.2);
\draw   [->, dashed] (-3.5+ 1*\j,1.8)-- (-3.5+ 1*\j + 1,1.2);
   \draw   [->, dashed] (-3.5+ 1*\j,1.8)-- (-3.5+ 1*\j - 1,1.2);
 }
 
  \foreach \j in {8} {
 \draw [fill= yellow](-3.5 + 1*\j, 2) circle (0.2);
  \draw [fill= yellow](-3.5 +1*\j, 0) circle (0.2);
\node[draw =none] (s2) at (-3.5+1*\j,1 ) {\footnotesize$+$};
\draw (-3.5 +1*\j, 1) circle (0.2);
 \draw   [->] (-3.5+1*\j,0.8)-- (-3.5+ 1*\j,0.2);
  \draw   [->] (-3.5+1*\j,1.8)-- (-3.5+ 1*\j,1.2);
 \draw   [->, dashed] (-3.5+ 1*\j,1.8)-- (-3.5+ 1*\j + 1,1.2);
 }
 
 \foreach \j in {6,14} {
 \draw [fill= yellow](-3.5 + 1*\j, 2) circle (0.2);
  \draw [fill= yellow](-3.5 +1*\j, 0) circle (0.2);
\node[draw =none] (s2) at (-3.5+1*\j,1 ) {\footnotesize$+$};
\draw (-3.5 +1*\j, 1) circle (0.2);
 \draw   [->] (-3.5+1*\j,0.8)-- (-3.5+ 1*\j,0.2);
  \draw   [->] (-3.5+1*\j,1.8)-- (-3.5+ 1*\j,1.2);
 \draw   [->, dashed] (-3.5+ 1*\j,1.8)-- (-3.5+ 1*\j - 1,1.2);
 }
 

\node[draw =none] (s2) at (-3.5+ 1-2,-0.5 ) {\footnotesize$0$};
\node[draw =none] (s2) at (-3.5+ 2-2,-0.5 ) {\footnotesize$1$};
\node[draw =none] (s2) at (-3.5+ 3-2,-0.5 ) {\footnotesize$2$};
\node[draw =none] (s2) at (-3.5+ 4-2,-0.5 ) {\footnotesize$3$};
\node[draw =none] (s2) at (-3.5+ 3,-0.5 ) {\footnotesize$4$};
\node[draw =none] (s2) at (-3.5+ 4,-0.5 ) {\footnotesize$5$};
\node[draw =none] (s2) at (-3.5+ 5,-0.5 ) {\footnotesize$6$};
\node[draw =none] (s2) at (-3.5+ 6,-0.5 ) {\footnotesize$7$};
\node[draw =none] (s2) at (-3.5+ 7,-0.5 ) {\footnotesize$8$};
\node[draw =none] (s2) at (-3.5+ 8,-0.5 ) {\footnotesize$9$};
\node[draw =none] (s2) at (-3.5+ 9,-0.5 ) {\footnotesize$10$};
\node[draw =none] (s2) at (-3.5+ 10,-0.5 ) {\footnotesize$11$};
\node[draw =none] (s2) at (-3.5+ 11,-0.5 ) {\footnotesize$12$};
\node[draw =none] (s2) at (-3.5+ 12,-0.5 ) {\footnotesize$13$};
\node[draw =none] (s2) at (-3.5+ 13,-0.5 ) {\footnotesize$14$};
\node[draw =none] (s2) at (-3.5+ 14,-0.5 ) {\footnotesize$15$};
\node[draw =none] (s2) at (-3.5+ 15,-0.5 ) {\footnotesize$16$};

\foreach \i in {0,1}{
\draw[->,very thick, yellow](-3.5 + 2 +0.2-0.03-2+8*\i, 0) --(-3.5 + 2 + 1-0.2+0.03-2+8*\i, 0) ;
\draw[<-,very thick, blue](-3.5 + 2 +0.2-0.03-2+8*\i, 2) --(-3.5 + 2 + 1-0.2+0.03-2+8*\i, 2) ;

\draw[->,very thick, blue](-3.5 + 3 +0.2-0.03-2+8*\i, 2) --(-3.5 + 3 + 1-0.2+0.03-2+8*\i, 2) ;
\draw[<-,very thick, yellow](-3.5 + 3 +0.2-0.03-2+8*\i, 0) --(-3.5 + 3 + 1-0.2+0.03-2+8*\i, 0) ;

\draw[->,very thick, yellow](-3.5 + 4 +0.2-0.03-2+8*\i, 0) --(-3.5 + 4 + 1-0.2+0.03-2+8*\i, 0) ;
\draw[<-, very thick, blue](-3.5 + 4 +0.2-0.03-2+8*\i, 2) --(-3.5 + 4 + 1-0.2+0.03-2+8*\i, 2) ;

\draw[->,very thick, blue](-3.5 + 5+0.2-0.03-2+8*\i, 2) --(-3.5 + 5 + 1-0.2+0.03-2+8*\i, 2) ;
\draw[<-, very thick, yellow](-3.5 + 5 +0.2-0.03-2+8*\i, 0) --(-3.5 + 5 + 1-0.2+0.03-2+8*\i, 0) ;

\draw[->,very thick, yellow](-3.5 + 6 +0.2-0.03-2+8*\i, 0) --(-3.5 + 6 + 1-0.2+0.03-2+8*\i, 0) ;
\draw[<-,very thick, blue](-3.5 + 6 +0.2-0.03-2+8*\i, 2) --(-3.5 + 6 + 1-0.2+0.03-2+8*\i, 2) ;

\draw[<-,very thick, yellow](-3.5 + 7 +0.2-0.03-2+8*\i, 0) --(-3.5 + 7 + 1-0.2+0.03-2+8*\i, 0) ;
\draw[->,very thick, blue](-3.5 + 7 +0.2-0.03-2+8*\i, 2) --(-3.5 + 7 + 1-0.2+0.03-2+8*\i, 2) ;
}
\draw [green, very thick](-3.5 +3, 0) circle (0.25);
\draw [green, very thick](-3.5 +11, 0) circle (0.25);
\draw [green, very thick](-3.5 +3, 2) circle (0.25);
\draw [green, very thick](-3.5 +11, 2) circle (0.25);

\foreach \i in {0,1}{
\draw[->,very thick, blue](-3.5 + 6 +0.2-0.03-4+8*\i, 1.9-2-0.05) --(-3.5 + 6 + 1-0.2+0.03-4+8*\i, 1.9-2-0.05) ;
\draw[->,very thick, blue](-3.5 + 5 +0.2-0.03-4 +8*\i, 1.9-2-0.05) --(-3.5 + 5+ 1-0.2+0.03-4+8*\i, 1.9-2-0.05) ;

\draw[<-,very thick, blue](-3.5 + 6 +0.2-0.03-4+8*\i, 2.1-2+0.05) --(-3.5 + 6 + 1-0.2+0.03-4+8*\i, 2.1-2+0.05) ;
\draw[<-,very thick, blue](-3.5 + 6 +0.2-0.03-1-4+8*\i, 2.1-2+0.05) --(-3.5 + 6 + 1-0.2+0.03-1-4+8*\i, 2.1-2+0.05) ;

\draw[<-,very thick, blue](-3.5 + 8 +0.2-0.03-4+8*\i, 1.9-2-0.05) --(-3.5 + 8 + 1-0.2+0.03-4+8*\i, 1.9-2-0.05) ;
\draw[<-,very thick, blue](-3.5 + 7 +0.2-0.03-4+8*\i , 1.9-2-0.05) --(-3.5 + 7+ 1-0.2+0.03-4+8*\i, 1.9-2-0.05) ;

\draw[->,very thick, blue](-3.5 + 8 +0.2-0.03-4+8*\i, 2.1-2+0.05) --(-3.5 +8 + 1-0.2+0.03-4+8*\i, 2.1-2+0.05) ;
\draw[->,very thick, blue](-3.5 + 8 +0.2-0.03-1-4+8*\i, 2.1-2+0.05) --(-3.5 + 8 + 1-0.2+0.03-1-4+8*\i, 2.1-2+0.05) ;
}
\end{tikzpicture}
 \caption{Illustration of message assignment and cooperation  in Wyner's symmetric network. }
\label{fig-sym}
\vspace{-.5cm} 
\end{figure}

Each network is decomposed into three subsets of Tx/Rx pairs, $\mathcal T_{\text{silent}}$, $\mathcal T_{\text{fast}}$ and $\mathcal T_{\text{slow}}$, where 
\begin{itemize}
\item Txs in $\mathcal T_{\text{silent}}$ are silenced and Rxs  in $\mathcal T_{\text{silent}}$ do not take any action.

\item Txs in $\mathcal T_{\text{fast}}$ send only ``fast'' messages. The corresponding Txs/Rxs   are  called ``fast". 
\item Txs in $\mathcal T_{\text{slow}}$ send only ``slow'' messages. The corresponding Txs/Rxs   are  called ``slow".

\end{itemize}
We choose the sets $\mathcal T_{\text{silent}}$, $\mathcal T_{\text{fast}}$ and $\mathcal T_{\text{slow}}$ in a way that:
\begin{itemize}
\item the signals sent by the ``fast" Txs  do not interfere; and
\item silencing the Txs in $\mathcal T_{\text{silent}}$ decomposes the network into non-interfering subnets such that in each subnet there is a dedicated Rx,  called \emph{master Rx}, that  can send a  cooperation message to any other ``slow" Rx  in the same subnet in at most  $\Drss$ cooperation rounds.
\end{itemize}
 For example, consider Wyner's symmetric model (described in detail in Section~\ref{sec:sym}) where Txs and Rxs are aligned on a grid and cooperation is possible only between neighbouring Txs or Rxs. Interference at a given Rx is only from adjacent Txs. The network is illustrated in Figure~\ref{fig-sym}. This figure also shows a possible decomposition of the Tx/Rx pairs into the  sets $\mathcal T_{\text{silent}}$ (in white), $\mathcal T_{\text{fast}}$ (in yellow) and $\mathcal T_{\text{slow}}$ (in blue) when $\D = 6$. The proposed decomposition creates subnets with $7$ active Tx/Rx pairs where the Rx in the center of any subnet (e.g. Rx~$4$ in the first subnet) can serve as a master Rx as it   reaches any slow (blue) Rx in the same subnet in at most $\lfloor \Dr-1/2\rfloor = \lfloor (\D-2)/2\rfloor = 2$ cooperation rounds.  As required, transmissions from fast (yellow) Txs are only interfered by  transmissions from slow (blue) Txs.

\subsubsection{Precanceling  of ``slow'' interference at ``fast''  Txs}  Any ``slow" Tx ${k'}$ quantizes its pre-computed input signal $X_{k'}^n$  (how this signal is generated will be described under item 5)) and describes  the quantised signal $\hat{X}_{k'}^n$ during the last Tx-cooperation round  to all its neighbouring ``fast" Txs, \mw{which then precancel this interference on their transmit signals.} (Here, there is only a single Tx-cooperation round, but this item will be reused in later subsections where $\Dt>1$.) Fig.~\ref{fig-sym} illustrates the sharing of the described quantization information  with neighbouring ``fast" Txs for Wyner's symmetric model.

To describe this formally, for each $k\in\{1,\ldots, K\}$, we define  the \emph{ ``slow" interfering set} 
\begin{equation}\label{eq:Ik_slow}
\mathcal{I}_{\tk}^{(S)}\triangleq  \mathcal{I}_{\tk} \cap \mathcal T_{\text{slow}}.
\end{equation}
Also, we  denote by $\vect U_{k}^n(M_{k}^{(F)})$  the  non-precoded input signal precomputed at a given ``fast" Tx~$k$. (The following item 3) explains how to obtain  $\vect U_{k}^n(M_{k}^{(F)})$.) Tx~$k$   sends the inputs  
 \begin{equation} \label{eq:13}
\vect X_k^n= \vect U_{k}^n(M_{k}^{(F)}) - \sum_{ k '\in \mathcal I_{\tk}^{(S)}} \mathsf H_{k,k}^{-1}\mathsf H_{ k',k} \hat{\vect X}_{ k'}^n,
\end{equation} 
 over the channel. 
Since each  ``fast" Rx~$k$ is  not interfered by the signal sent at any other ``fast" Tx,
the  precoding in \eqref{eq:13} makes that a ``fast" Rx~$k$ observes the almost interference-free signal
\begin{equation} \label{eq:14}
\vect Y_{\tk}^n = \mathsf H_{k,k} \vect U_k^n + \underbrace{\sum_{ k' \in \mathcal I_{k}^{(S)}} \mathsf{H}_{ k' ,k} (\vect X_{ k'}^n - \hat{\vect X}_{ k'}^n) + \vect Z_{k}^n}_{\textnormal{disturbance}},
\end{equation}
where  the variance of above disturbance is around noise level and does not grow with $\P$.

 \subsubsection{Transmission of   ``fast'' messages} Each  ``fast" Tx~$k$ encodes its desired message $M_k^{(F)}$  using a codeword $\mathbf{U}_k^{(n)}(M_k^{(F)})$ from a  Gaussian point-to-point code of power $\P$. The corresponding Rx~$\tk$ applies a standard point-to-point decoding rule to directly decode this ``fast'' codeword  without Rx-cooperation   from its ``almost" interference-free outputs $\mathbf{Y}_k$, see \eqref{eq:14}. 

  \subsubsection{Canceling  ``fast'' interference at ``slow'' Rxs}
According to the previous item 3),  all ``fast" messages are decoded directly from the outputs without any Rx-cooperation. During the first Rx-cooperation round, all ``fast" Rxs can thus share their decoded messages  with all their neighbouring ``slow" Rxs, which can cancel the  corresponding interference  from their receive signals. More formally, we define the \emph{fast interference set}
\begin{equation}\label{eq:IkF}
\mathcal{I}_{\tk}^{(F)}\triangleq \mathcal{I}_{\tk} \cap  \mathcal{T}_{\text{fast}}
\end{equation} as   the set of ``fast" Txs   whose signals interfere at Rx $\tk$. Each ``slow" Rx $\tk$  forms the new signal
 \begin{equation} \label{eq:15}
  \hat{\vect Y}_{\tk}^n: = \vect Y_{\tk}^n -\sum_{\hat k \in \mathcal{I}_{\tk}^{(F)}} \mathsf H_{\hat k,k}  \vect X_{\hat k}^n (\hat M_{\hat k}^{(F)}),
 \end{equation}
and decodes its desired ``slow" message based on this new signal following the steps  described in the following item 5).
 Fig.~\ref{fig-sym} illustrates with yellow arrows the sharing of  decoded ``fast'' messages with neighbouring ``slow" Rxs in Wyner's symmetric model.

 \subsubsection{Transmission and reception of ``slow'' messages using CoMP reception}
  Each ``slow" Tx~$k$ encodes its message $M_k^{(S)}$ using a codeword $\vect X_k^n(M_{k}^{(S)})$ from  a Gaussian point-to-point code of power~$\P$. ``Slow" messages are decoded based on the new outputs $\hat{\vect Y}_{\tk}^n$ in \eqref{eq:15}. CoMP reception  is employed to decode all ``slow" messages in a given subnet. That means, each ``slow" Rx~$\tk$ applies a rate-$\frac{\L}{2} \log (1 + \P)$ quantizer to the new output 
 signal  $\hat{\vect Y}_k^n$, and sends the quantization information over the cooperation links to the {master} Rx in its subnet. Each master Rx reconstructs all the quantized signals and  jointly decodes the ``slow'' messages, before sending them back to their intended Rxs.  By  item 4)   the influence of ``fast" transmissions has been canceled on the ``slow" receive signals. 

\subsubsection{MG Analysis}
In the described scheme, all transmitted ``fast" and  ``slow'' messages can be sent reliably  at MG $\L$ because all interference is cancelled (up to noise level) either at the Tx or the Rx side, and because Txs and Rxs are equipped with $\L$ antennas each.

 The presented coding scheme thus achieves the MG pair
\begin{equation} \label{BOTH}
\left ( \S^{(F)} = \S^{(F)}_{\text {both}}, \; \;  \S^{(S)} =\S^{(S)}_{\text {both}} \right ),
\end{equation}
where 
\begin{equation} \label{eq:MG}
\S^{(F)}_{\text {both}} \triangleq \L \cdot  \varlimsup_{K\to \infty} \frac{|\mathcal T_{\text{fast}}|}{K}  \quad \text{and} \quad \S^{(S)}_{\text {both}} \triangleq \L \cdot  \varlimsup_{K\to \infty}  \frac{|\mathcal T_{\text{slow}}|}{K} .
\end{equation}

 
The scheme  we described so far requires  different   cooperation rates on the various  Tx- or Rx-cooperation links. To evenly balance the load on the  Tx-cooperation links and on the  Rx-cooperation links, different versions of the scheme with different choices of the sets $\mathcal T_{\text{silent}}$, $\mathcal T_{\text{fast}}$ and $\mathcal T_{\text{slow}}$ and different cooperation routes can be time-shared.   The main quantity of interest is then the \emph{average cooperation load}, which for the scheme above is characterized as follows. During the single Tx-cooperation round, each ``fast" Tx $k$ receives a quantised version of the transmit signal of each of its ``slow" interferers  $\hat k\in\mathcal{I}_k^{(S)}$. Since each quantisation message is of prelog $\L$, the average required Tx-cooperation prelog equals
\begin{equation} \label{mtbone}
\mtbone \triangleq \L \cdot \varlimsup_{K\to \infty}\frac{\sum_{k \in \mathcal T_{\text{fast}}} |\mathcal I_k^{(S)}|}{\q_{K,\Tx}},
\end{equation}
where $\q_{K,\Tx}$ denotes the total number of  Tx-cooperation links in the network.

There are three types of Rx-cooperation messages. In  the first Rx-cooperation round, each ``slow" Rx $k$ obtains a  decoded message from each of its ``fast" interferers $\hat k\in \mathcal{I}_k^{(F)}$. The total number of messages sent in this first round is thus $\sum_{k \in \mathcal T_{\text{slow}}}|\mathcal I_k^{(F)}|$ and each is of prelog $\L$.  In Rx-cooperation rounds $2,\ldots, \lfloor \frac{\Dr-1}{2}\rfloor+1$, ``slow" Rxs send quantized versions of their output signals to the master Rx in the same network. Each of these messages is of prelog $\L$ and the total number of such messages equals $\sum_{k \in \mathcal T_{\text{slow}}}  \gamma_{\Rx, k}$, where $\gamma_{\Rx, k}$ denotes the number of cooperation rounds required for ``slow" Rx~$k$ to reach the master Rx in its subnet. In rounds $ \lfloor \frac{\Dr-1}{2}\rfloor+2,\ldots,  \Dr$, the master Rx sends the decoded messages to all the ``slow" Rxs in the subnet. Each of these messages is again of prelog $\L$ and the total number of such messages is again $\sum_{k \in \mathcal T_{\text{slow}}}  \gamma_{\Rx, k}$. To summarize, each of the transmitted messages is of prelog $\L$ and thus the  {average cooperation} prelog required per Rx-cooperation link is:
\begin{equation} \label{mrbone}
\mrbone \triangleq \L \cdot \varlimsup_{K\to \infty}  \frac{\sum_{k \in \mathcal T_{\text{slow}}} \big (|\mathcal I_k^{(F)}| +  2 \gamma_{\Rx, k} \big )}{\q_{K,\Rx}},
\end{equation}
where $\q_{K,\Rx}$ denotes the total number of  Rx-cooperation links in the network.

\begin{remark}\label{rem1}
If the master Rx of a subnet is a ``fast'' Rx, it does not have to send its decoded message to its ``slow'' neighbours, because it  decodes all ``slow'' messages jointly. In this case, less Rx-cooperation prelog is required.
\end{remark}


\subsection{Coding scheme to transmit both ``fast'' and ``slow'' messages with CoMP transmission:} \label{sub:both2}
This second scheme splits the total number of cooperation rounds $\D$ as:
\begin{equation}
\Dt=\D-1 \quad \textnormal{and} \quad \Dr= 1.
\end{equation} 
Similarly to the  previous Subsection \ref{sub:both}, the scheme is described by $5$ items:

\subsubsection{Creation of subnets and message assignment}
This item is similar to item 1) of Subsection \ref{sub:both}, but  
 the sets $\mathcal T_{\text{silent}}$, $\mathcal T_{\text{fast}}$ and $\mathcal T_{\text{slow}}$ are chosen in a way that:
\begin{itemize}
\item as before,  the signals sent by the ``fast" Txs  do not interfere; and
\item silencing the Txs in $\mathcal T_{\text{silent}}$ decomposes the network into non-interfering subnets  so that  in each subnet there is a dedicated \emph{master Tx} that can send a  cooperation message   to any other ``slow" Tx  in the same subnet  in at most  $\Dtss$ cooperation rounds.
\end{itemize}


Items 2)-4) remain as described in  Subsection \ref{sub:both}. Item 5) is replaced by the following item.
\emph{5) Transmission and reception of  ``slow'' messages using CoMP transmission:}
``Slow" messages are transmitted using standard CoMP transmission techniques that can ignore interference from ``fast" Txs (due to the post-processing in item 4)) but  account for the modified interference graph  and  the modified channel matrix between slow messages caused by the precanceling performed under item 2). 
The receivers decode based on the new outputs $\hat {\vect Y}_{\tk}^n$  in \eqref{eq:15}. 

We describe CoMP transmission in this context more formally. During the first $\lfloor \frac{\Dt-1}{2}\rfloor$ Tx-cooperation rounds, each  ``slow" Tx of a subnet,   sends its message to the master Tx of the subnet. This latter  encodes all received ``slow" messages using individual Gaussian codebooks and    precodes them  so as to cancel all the interference from other ``slow" messages at the corresponding Rxs. I.e., it produces signals so that when they are transmitted over the active antennas in the cell,  the signal observed at each ``slow" Rx only depends on the ``slow" message sent by the corresponding Tx but not on the other ``slow" messages. 
The master Tx  applies a Gaussian vector quantizer on these precoded signals and sends the  quantization information over the cooperation links  to the corresponding Txs during the Tx-cooperation rounds $\lfloor \frac{\Dt-1}{2}\rfloor+1$ to $\Dt-1$. This is possible by the way we defined the master Txs. 
All ``slow" Txs reconstruct the quantized signals $\hat{\vect X}_{k}^n$ intended for them and send them over the network:
$\vect X_k^n \triangleq \hat{\vect X}_{k}^n$.

Each ``slow" Rx~$\tk$ decodes its desired  message  from the modified output sequence $\hat {\vect Y}_{\tk}^n$ defined in \eqref{eq:15} using a standard point-to-point decoder. 

\textit{Analysis:}
Similarly to Subsection~\ref{sub:both}, each transmitted message can be sent reliably at MG $\L$, and thus the  scheme achieves the  MG pair in \eqref{BOTH}. 

The load on the different cooperation links is again unevenly distributed across links, and thus, by time-sharing and symmetry arguments,  the average Rx- and Tx-cooperation rates are the limiting quantities. 
The required average Rx-cooperation rate is easily characterized as:
 \begin{IEEEeqnarray}{rCl}
 \mrbtwo &\triangleq &\L \cdot \varlimsup_{K\to \infty}  \frac{\sum_{k \in \mathcal T_{\text{slow}}} |\mathcal I_k^{(F)}|}{\q_{K, \Rx}}, \label{mrbtwo}
  \end{IEEEeqnarray}
  because  Rx-cooperation takes place in a single round, during which each ``slow" Rx $k$ learns all  decoded ``fast" messages that interfere their receive signals and these messages are of MG $\L$. 
To calculate the required average Tx-cooperation rate, define for each $k\in\mathcal{T}_{\text{slow}}$ the positive parameter $\gamma_{\Tx,k}$ to be the number of cooperation hops required from Tx~$k$ to reach the master Tx in its subnet. During the first $\lfloor \frac{\Dt-1}{2}\rfloor$ Tx-cooperation rounds, a total of $\sum_{k \in \mathcal T_{\text{slow}}} \gamma_{\Tx, k}$ cooperation messages of MG $\L$ are transmitted from the ``slow" Txs to the  master Txs in their subnet. The same number of Tx-cooperation messages, all of MG $\L$, is also conveyed during rounds $\lfloor \frac{\Dt-1}{2}\rfloor+1, \ldots, 2\lfloor \frac{\Dt-1}{2}\rfloor$, now from the master Tx to the ``slow" Txs in the subnet. During the last round, ``slow" Txs convey their messages to the adjacent ``fast" Txs that are interfered by their signals. Some of these signals, however have already been shared during Tx-cooperation rounds  $\lfloor \frac{\Dt-1}{2}\rfloor+1, \ldots, 2\lfloor \frac{\Dt-1}{2}\rfloor$, and thus do not have to be sent again. The total number of cooperation messages during the last Tx-cooperation rounds is thus only equal to $\sum_{k\in \mathcal{T}_{\text{fast}}} |\mathcal I_k^{(S)}| -q$, where $q$ denotes the number of the messages that have already been sent in previous rounds. We will chracterize the value of $q$ when we analyze specific networks. To summarize, the average required Tx-cooperation rate of our scheme is:
\begin{IEEEeqnarray}{rCl}  
\mtbtwo & \triangleq & \L \cdot \varlimsup_{K\to \infty}  \frac{ \sum_{k \in \mathcal T_{\text{slow}}} 2 \gamma_{\Tx, k} + \sum_{k \in \mathcal T_{\text{fast}}} |\mathcal I_k^{(S)}| -q }{\q_{K, \Tx}}. \label{mtbtwo}
 \end{IEEEeqnarray}

	\subsection{Coding scheme to transmit  only ``slow'' messages with CoMP reception and transmission:}	\label{sub:onlyslow}
	In principle,  since any ``fast'' message satisfies the constraints on ``slow'' messages,  we can use the schemes provided in Subsections \ref{sub:both} and \ref{sub:both2} to send only ``slow'' messages. Sometimes, the following scheme however performs better because it requires less  Tx- or Rx-cooperation rates. 
	Choose a set 
$\mathcal T_{\text{silent}}\subseteq [K]$ and  silence the  Txs in this set, which decomposes the network in non-interfering subnets. The remaining Txs in $\mathcal T_{\text{slow}}:=[K]\backslash \mathcal{T}_{\text{silent}}$ send only ``slow" messages using CoMP transmission or reception. The set $\mathcal{T}_{\text{silent}}$ thus has to be chosen such that in each subnet there is a dedicated  \emph{master Rx} (or \emph{master Tx}), which can be reached by any other Rx (Tx) in the subnet  in at most  $\Ds$ cooperation rounds. 
Both versions achieve the MG pair
	\begin{equation} \label{mgs}
	\left ( \S^{(F)} = 0, \; \;  \S^{(S)} = \S^{(S)}_{\max}  \right ),
	\end{equation}
	where 
\begin{equation} \label{ssm}
\S^{(S)}_{\max} \triangleq\L\cdot \varlimsup_{K\to \infty}\frac{|\mathcal T_{\text{slow}}|}{K}.
\end{equation}
The CoMP-reception scheme requires no Tx-cooperation  but  average Rx-cooperation prelog  
	\begin{equation} \label{mrstwo}
	\mrstwo \triangleq \L \cdot \varlimsup_{K\to \infty}  \frac{\sum_{k \in \mathcal T_{\text{slow}} }2\gamma_{\Rx,k}}{\q_{K, \Rx}},
	\end{equation} 
and the CoMP-transmission scheme   no Tx-cooperation but   average Tx-cooperation rate  
	\begin{equation} \label{mtsone}
	\mtsone \triangleq \L \cdot \varlimsup_{K\to \infty}  \frac{\sum_{k \in \mathcal T_{\text{slow}} }2\gamma_{\Tx,k}}{\q_{K, \Tx}},
	\end{equation} 
	where recall that  $\gamma_{\Rx,k} ,\gamma_{\Tx,k} \in \{1, \ldots, \Ds\}$ denote  the number of cooperation hops required from a Rx~$k$ or a Tx~$k$ to reach the master Rx or the master Tx in its subnet. 
%
%

\subsection{Coding scheme without cooperation:} \label{sub:onlyfast}
	Choose a set of Txs $\mathcal T_{\text{silent}} \subseteq [K]$  so that the remaining Txs $\mathcal{T}_{\textnormal{active}}:=[K]\backslash \mathcal{T}_{\text{silent}}$  do not interfere, and send  ``slow" or  ``fast" over the resulting interference-free links.  The scheme requires no cooperation and achieves for any $\beta \in [0, 1]$ the MG pair
	\begin{equation}\label{sfmax}
	\left ( \S^{(F)} = \beta \S_{\text{no-coop}}, \; \;  \S^{(S)} = (1-\beta)\S_{\text{no-coop}}  \right ), 
	\end{equation}
	where
	\begin{equation} \label{sfm}
		\S_{\text{no-coop}} \triangleq\L\cdot \varlimsup_{K\to \infty}\left (1- \frac{|\mathcal T_{\text{silent}}|}{K}\right ).
	\end{equation}

\section{Wyner's Symmetric Linear Model}\label{sec:sym}

Consider Wyner's symmetric linear cellular model where  cells are aligned in a single dimension and  signals of users that lie in a given cell interfere only with signals sent in the two adjacent cells. Since the focus of the paper is on the MG, for simplicity, we assume only a single mobile user in each cell, each wishing to communicate with the corresponding BS of the cell.  We shall further assume that the number of cells  $K$ and the maximum delay $\D$ are  even.

  The input-output relation of the network is   
\begin{equation}\label{Eqn:Channel}
\vect{Y}_{k,t} = \mathsf{H}_{k,k} \vect{X}_{k,t} + \mathsf{H}_{k-1,k} \vect{X}_{k-1,t}+ \mathsf{H}_{k+1,k} \vect{X}_{k+1,t} +\vect{Z}_{k,t},
\end{equation}
where $\vect{X}_{0,t} = \vect{0}$ for all $t$, 
and the interference set at a given user  $k$ is
\begin{equation}
\mathcal{I}_{k}=\{k-1,k+1\},
\end{equation}where indices out of the range $[K]$ should be ignored. 
In this model, Rxs and Txs can cooperate with the two Rxs and Txs  in the adjacent cells, so 
\begin{equation}
\Ntk =\{k-1,k+1\}\quad \textnormal{and} \quad  \Nrk =\{k-1,k+1\}.
\end{equation}
Fig.~ \ref{fig3-t2} illustrates the interference pattern of the network and the available cooperation links. As can be seen from this figure,  Txs~$1$ and~$K$  and  Rxs~$1$ and~$K$ have a single outgoing cooperation link and all other Txs and Rxs in this network have two outgoing cooperation links. Thus, the total numbers of Tx- and of Rx-cooperation links  both are   
\begin{equation}
\label{eq:CoopLink}
\q_{K, \Tx} = \q_{K, \Rx} = 2K-2.
\end{equation}

\subsection{Choice of Tx/Rx Sets for the Schemes in Section~\ref{coding}} \label{sub:IV-A}
\subsubsection{``Fast'' and ``slow'' messages with CoMP reception} \label{sub:symbothR} 
\mw{For the mixed-delay scheme, choose the Tx/Rx set association
  in Fig.~\ref{fig-sym},} where ``fast'' Tx/Rx pairs are in yellow, ``slow''  in blue, and silenced in white. I.e., set
\begin{subequations}\label{eq:set}
\begin{IEEEeqnarray}{rCl}
\mathcal{T}_{\text{silent}}& = & \left \{ \ell \left (\D+2 \right ) \colon \ell=1,\ldots, \left \lfloor \frac{K}{\D+2} \right \rfloor \right \} ,\\
\mathcal{T}_{\text{fast}}& = &\{1,3,\ldots, K-1\},\\
\mathcal{T}_{\text{slow}}& = &\{1,\ldots,  K\} \backslash \{\mathcal{T}_{\text{silent}}, \mathcal{T}_{\text{fast}}\}.
\end{IEEEeqnarray}
\end{subequations}
For this choice, 
transmissions of ``fast" messages are interfered only by transmissions of ``slow" messages and 
for any $\ell$, the Tx/Rx pairs in 
\begin{equation}
\mathcal T_{\ell} \triangleq \{\ell(\D+2)+1,\ldots, (\ell+1)(\D+2)-1 \}
\end{equation} form a subnet for which Rx~$\ell(\D+2)+\D/2+1$   can act as  the master Rx because it can be reached by any \emph{``slow'' Rx (i.e., even Rx)}  in its subnet in at most $(\Dr-1)/2$ cooperation hops.

By \eqref{eq:MG} and \eqref{eq:set}, the scheme achieves the MG pair $(\S^{(F)}  = \sfbw, \S^{(S)} = \ssbw)$  
where 
\begin{equation} \label{both1}
\sfbw \triangleq \frac{\L}{2} \quad \textnormal{and} \quad \ssbw \triangleq \L \cdot \frac{\D}{2(\D + 2)}.
\end{equation}

To analyze the required cooperation prelogs of the  scheme, $\mtbonew$ and $\mrbonew$, we evaluate the formulas in \eqref{mtbone} and \eqref{mrbone}.  
 We  have for each subnet $\ell\in\{1,\ldots, \lfloor K/(\D+2)\rfloor\}$:
\begin{equation}\label{eq:sum0}
\sum_{k \in \mathcal{T}_{\text{fast}} \cap \mathcal{T}_{\ell}} |\mathcal{I}_{k}^{(S)}| = 2+2(\D/2-1)=\D.
\end{equation}
In the limit $K\to \infty$,  we obtain
\begin{equation}\label{eq:mtbonew}
\mtbonew  = \L \cdot \frac{\D}{2(\D+2)}.
\end{equation}

To calculate the required Rx-cooperation prelog $\mrbonew$, notice that 
$|\mathcal{I}_k^{(F)}|=2$. Since there are $\D/2$ ``slow" Rxs in each subnet $\mathcal{T}_{\ell}$:
	\begin{equation}\label{eq:sum1}
	\sum_{k \in \mathcal{T}_{\text{slow}} \cap \mathcal{T}_{\ell}} |\mathcal{I}_{k}^{(F)}| = 2\cdot \frac{\D}{2}=\D.
	\end{equation}
 
  In addition, Rxs also exchange cooperation messages to enable CoMP reception. Thereby,  the quantization message produced by a ``slow" Rx $k=\ell(\D+2)+i$, for $i \in\{2,4,\ldots, \D-2\}$,  has to  propagate over  $\gamma_{\Rx,k}=|\D/2+1-i|$ hops to reach the subnet's master Rx. If $\D/2 + 1$ is even, 
  	\begin{equation}\label{eq:sum2}
  	\sum_{k \in \mathcal{T}_{\text{slow}} \cap \mathcal{T}_{\ell}} \gamma_{\Rx,k}= \sum_{i\in\{2,4,\ldots, \D/2-1\}}2 \cdot ( \D/2+1-i) = \frac{1}{2} \left( \frac{\D^2}{4} - 1\right).
  	\end{equation}
 Then, according to \eqref{mrbone}, \eqref{eq:CoopLink}, \eqref{eq:sum1}, and \eqref{eq:sum2}, when $\D/2+1$ is even,  in the limit as $K \to \infty$: 
\begin{equation}\label{eq:mrbonew}
\mrbonew  = \L \cdot \frac{\D + \frac{\D^2}{4}-1 }{2(\D + 2)},  \quad \text{for} \;\; \D/2 + 1 \;\; \text{even}.
\end{equation}
	When $\D/2 +1$ is odd, the sum in \eqref{eq:sum2} evaluates to $\frac{\D^2}{8}$. Moreover, in this case, the master Rx is a ``fast" Rx. It does not have to send its decoded message to any neighbour, as it locally decodes all ``slow" messages of the subnet. So, (see also Remark~\ref{rem1}), the nominator in \eqref{mrbone} can be reduced by $2$. Putting all these together,
	  we obtain $\mrbonew  = \L \cdot \frac{\D + \frac{\D^2}{4}-2 }{2(\D + 2)}$ when $\D/2+1$ is odd.

\subsubsection{``Fast'' and ``slow'' messages with CoMP transmission} \label{sub:symbothT}
Choose the same  cell association  as for the CoMP reception scheme described in \eqref{eq:set} and depicted in Fig.~\ref{fig-sym}.  Under this cell association, Tx~$\D/2 + 1$  can act as a master Tx because it can be reached by any ``slow" (even) Tx in its subnet in at most $(\Dt-1)/2$ cooperation rounds. 
Since the same cell partitioning is used, namely \eqref{eq:set}, this scheme achieves the same MG pair as with CoMP reception, see  \eqref{both1}. Moreover, by \eqref{mrbtwo} and \eqref{eq:sum1} in the limit as $K\to \infty$,  the required average Rx-cooperation prelog is 
\begin{equation}\label{eq:mrbtwow}
\mrbtwow= \L \cdot \frac{\D}{2(\D+2)}. 
\end{equation}
Similarly, consider  \eqref{mtbtwo} and \eqref{eq:sum0}  and notice that for $\D/2+1$ even,
  	\begin{equation}\label{eq:sum2b}
  	\sum_{k \in \mathcal{T}_{\text{slow}} \cap \mathcal{T}_{\ell}} \gamma_{\Tx,k}= \sum_{i\in\{2,4,\ldots, \D/2-1\}}2( \D/2+1-i) = \frac{1}{2} \left( \frac{\D^2}{4} - 1\right), 
  	\end{equation}
  	whereas for $\D/2 +1$ odd, this sum evaluates to $\frac{\D^2}{8}$.
We consider  the $q$-term in \eqref{mtbtwo},  which characterizes the number of quantization messages describing the ``slow" signals that are counted twice: once for the CoMP transmission and once for the interference mitigation at ``fast" transmitters. In each subnet, $\D/2-1$ such messages are double-counted, when $\D/2+1$ is even, and $\D/2$ messages are double-counted when $\D/2+1$ is odd. 
Therefore, and  according to \eqref{mtbtwo}, \eqref{eq:sum0}, \eqref{eq:sum2b}, 
when $K \to \infty$, 
the average Tx-cooperation prelog required by the scheme is 
 \begin{IEEEeqnarray}{rCl}\label{eq:mtbtwow}
  \mtbtwow &=&\L\cdot   \frac{\D + \frac{\D^2}{4}-1 -\D/2+1 }{2(\D+2)}= \L \cdot \frac{\D}{8},
 \end{IEEEeqnarray}
irrespective of whether  $\D/2+1$ is even or odd.


\subsubsection{Transmitting only ``slow'' messages with CoMP reception and transmission} \label{sub:symslow}
Consider the scheme  in Subsection \ref{sub:onlyslow} that transmits only ``slow" messages, either using  CoMP transmission  or CoMP reception. For both schemes we regularly silence every $\D+2$nd Tx, i.e., as in   the two previous  subsections, $\mathcal{T}_{\textnormal{silent}} \triangleq \big\{\ell (D+2)\colon \ell=1, \ldots, \lfloor \frac{K}{\D+2}\rfloor \big\}$. Also, we set $\mathcal{T}_{\textnormal{slow}} = [K] \backslash \mathcal{T}_{\textnormal{silent}}$. These choices are permissible, because all Txs (or Rxs) in a subnet $\mathcal{T}_{\ell}=\{(\ell-1) (\D+2)+1,\ldots, \ell (\D+2)-1 \}$ can reach the subnet's  central Tx~$(\ell-1)(\D+2)+\D+1$ (or Rx~$(\ell-1)(\D+2)+\D+1$) in at most $\D/2$ cooperation hops.

 By \eqref{ssm}, the scheme in Subsection~\ref{sub:onlyslow} achieves the MG pair $( \S^{(F)} = 0, \; \;  \S^{(S)} = \ssmw)$ where 
\begin{equation}\label{slow-sym2}
\ssmw \triangleq \L \cdot \frac{\D + 1}{\D+ 2} .
\end{equation}
With CoMP reception, this scheme does not use any Tx-cooperation. To calculate the  Rx-cooperation prelog, we use the fact that Rx~ $k=\ell(\D+2)+i$,  for positive integers $\ell$ and $i\leq \D+1$, reaches the master Rx in its subnet in $\gamma_{\Rx,k}=|\D/2+1-i|$ hops. Since:
\begin{equation}
2\sum_{k \in  \mathcal{T}_{\ell}} \gamma_{\Rx,k} = 4 \sum_{i =1}^{\D/2} i= \frac{\D(\D+2)}{2},
\end{equation}
  by \eqref{mrstwo}, in the limit as $K \to \infty$, the average Rx-cooperation prelog tends to
 \begin{IEEEeqnarray}{rCl}\label{slow-mu}
  \mrstwow &=&  \L \cdot \frac{\D}{4}.
 \end{IEEEeqnarray}
Similar conclusions show that when CoMP transmission is used instead of CoMP reception,  the scheme requires zero Rx-cooperation prelog and a Tx-cooperation prelog of $\mtsonew = \mrstwow$.


\subsubsection{No-cooperation scheme} \label{sub:symfast}
Consider the no-cooperation scheme in Subsection~\ref{sub:onlyfast}. For Wyner's symmetric network we  create   non-interfering point-to-point links by silencing all even Txs in the network, i.e.,  by  choosing $\mathcal T_{\text{silent}} \triangleq \{2,4, \ldots 2\lfloor \frac{K}{2} \rfloor \}$. Since all odd receivers remain active, the sum-prelog in \eqref{sfm}  for this network evaluates to
\begin{equation} \label{fast-sym} 
\sfmw \triangleq \frac{\L}{2}.
\end{equation}

\subsection{Achievable MG Regions}
Recall the definitions of $\sfbw$, $ \ssbw$, $\ssmw$, $\ssncw$  in \eqref{both1}, \eqref{slow-sym2}, and \eqref{fast-sym} and the definitions of $\mtbonew, \mrbonew, \mrbtwow, \mtbtwow$ in \eqref{eq:mtbonew}, \eqref{eq:mrbonew}, \eqref{eq:mtbtwow}, and  \eqref{eq:mrbtwow}. Define further
\begin{IEEEeqnarray}{rCl}\label{eq:alpha}
\alpha &\triangleq& \max\left\{   \min\left\{ \frac{\muT}{\mtbonew}, \frac{\muR}{\mrbonew}\right\},\;  \min\left\{ \frac{\muT}{\mtbtwow}, \frac{\muR}{\mrbtwow}\right\} \right\}.
\end{IEEEeqnarray}
and
\begin{IEEEeqnarray}{rCl}
\S^{(S)}_{\text{sym},1}(\alpha) \triangleq \alpha \ssmw + (1 - \alpha)\ssncw\qquad& \\
  \S^{(F)}_{\text{sym},2}(\alpha) \triangleq \alpha \sfbw + (1-\alpha)\sfmw \qquad &\textnormal{and}& \qquad 
 \S^{(S)}_{\text{sym},2}(\alpha)\triangleq \alpha \ssbw\\
  \S^{(F)}_{\text{sym},3}(\alpha) \triangleq \alpha \sfbw  \qquad &\textnormal{and}& \qquad   \S^{(S)}_{\text{sym},3}(\alpha)\triangleq \alpha \ssbw + (1- \alpha) \ssmw.  \IEEEeqnarraynumspace
\end{IEEEeqnarray}
According to the arguments in the previous subsection, the following regions of MG pairs are achievable depending  on the available cooperation prelogs $\muT$ and $\muR$.

 \begin{theorem} [Achievable MG Region: Wyner's Symmetric Model] \label{lemma1} Assume $\D \ge 2$ \maw{and even}.

 When $\muR \geq \mrbonew$ and  $\muT \geq \mtbonew$; or when $\muR \geq \mrbtwow$ and  $\muT \geq \mtbtwow$:
 		\begin{IEEEeqnarray}{rCl}\label{eq:trapezoidal1}
 		\textnormal{convex hull}\Big(  (0,0), \ (0, \ssmw), \  (\sfbw, \ssbw),  \  (\sfmw, 0) \Big) \subseteq \mathcal{S}^\star(\muT,\muR,\D).\IEEEeqnarraynumspace
 		\end{IEEEeqnarray}


 	 When  $\muR \ge \mrstwow$ and $\muT < \mtbonew$; or when  $\muT \ge \mtsonew$ and $\muR < \mrbtwow$: 
 		\begin{IEEEeqnarray}{rCl}\label{eq:trapezoidal2}
 	\lefteqn{	\textnormal{convex hull}\Big(  (0,0), \ (0, \ssmw),\  (\S^{(F)}_{\text{sym},3}(\alpha),  \S^{(S)}_{\text{sym},3}(\alpha)),}\nonumber \hspace{4cm}\\
 	 &&\hspace{0cm} \ (\S^{(F)}_{\text{sym},2}(\alpha), \S^{(S)}_{\text{sym},2}(\alpha)),  \  (\sfmw, 0) \Big) \subseteq \mathcal{S}^\star(\muT,\muR,\D). 
 		\end{IEEEeqnarray}
 		
 When $\muR < \mrbonew$  or when $\muT < \mtbtwow$: 
 		\begin{IEEEeqnarray}{rCl}\label{eq:trapezoidal3}
 		\textnormal{convex hull}\Big(  (0,0), \ (0, \S^{(S)}_{\text{sym},1}(\alpha)), \  (\S^{(F)}_{\text{sym},2}(\alpha), \S^{(S)}_{\text{sym},2}(\alpha)),  \  (\sfmw, 0) \Big)\subseteq \mathcal{S}^\star(\muT,\muR,\D). \IEEEeqnarraynumspace
 		\end{IEEEeqnarray}

 \end{theorem}

\begin{proposition}[Outer Bound on the MG Region: Wyner's Symmetric  Model] \label{conv:sym}
Any MG pair $(\S^{(F)}, \S^{(S)})$ in $\mathcal{S}^\star(\muT,\muR,\D)$ satisfies 
\begin{IEEEeqnarray}{rCl}
\S^{(F)} &\le& \frac{\L}{2},\\
\S^{(F)} + \S^{(S)} &\le& \L \cdot \frac{\D+1}{\D+2}.
\end{IEEEeqnarray}
 \end{proposition}
  \begin{IEEEproof}
 Specialize the \emph{MAC-Lemma for interference networks with cooperation}  \cite[Lemma 1]{Wiggeretal} to
	\begin{IEEEeqnarray}{rCl}
	\mathcal J_{\text{outputs}} &\triangleq& \bigcup_{\ell \in \{1, \ldots, \lceil \frac{K}{2(\D+2)}\rceil \}}  \left \{2+(\ell-1)(2\D+4), \ldots, \ell(2\D+4)-1 \right \}, \\
	 	\mathcal J_{\text{inputs}} &\triangleq& \bigcup_{\ell \in \{1, \ldots, \lceil \frac{K}{2(\D+2)}\rceil \}}  \left \{\D+2+(\ell-1)(2\D+4), \ldots, \D+3+ (\ell-1)(2\D+4) \right \}, \IEEEeqnarraynumspace \\
 	\mathcal J_{\text{messages}} &\triangleq& \bigcup_{\ell \in \{1, \ldots, \lceil \frac{K}{2(\D+2)}\rceil \}}  \left \{\D+2-\Dt+(\ell-1)(2\D+4), \ldots, \D+3+\Dt + (\ell-1)(2\D+4) \right \}. \nonumber \\[-3ex]&& \hspace{13cm}
\blacksquare
 	\end{IEEEeqnarray}
	 \end{IEEEproof}
 
 \begin{corollary}\label{cor1}
If
 \begin{equation}
\big( \muR \ge \mrbonew \quad \text{and} \quad \muT \ge \mtbonew \big) \hspace{.6cm} \text{or} \hspace{.6cm} \big( \muR \ge \mrbtwow \quad \text{and} \quad \muT \ge \mtbtwow \big),
 \end{equation}
the optimal MG region $\mathcal{S}^\star(\muT,\muR,\D)$ coincides with the trapezoid in \eqref{eq:trapezoidal1}. 
 \end{corollary}
  \begin{IEEEproof}
 	Follows directly by Theorem~\ref{lemma1} and  Proposition~\ref{conv:sym}.
 \end{IEEEproof}
By Corollary~\ref{cor1},   for large cooperation prelogs $\muT$ and $\muR$,  imposing a stringent delay constraint on the ``fast'' messages never penalizes the maximum achievable sum-MG of the system: the same sum-MG can be achieved as if only ``slow" messages were sent.

The next corollary characterizes the optimal MG region $\mathcal{S}^\star(\muT,\muR,\D)$ when one of the two cooperation prelogs ($\mu_{\Tx}$ or $\mu_{\Rx}$) is small and the other large, and when $\S^{(F)}$ lies below a certain threshold. The corollary shows that also in this regime the  same maximum sum-MG can be achieved as if only ``slow" messages were sent.  When $\S^{(F)}$ exceeds this threshold, our achievable MG region in \eqref{eq:trapezoidal2} shows a penalty in sum-MG which increases linearly with the ``fast" MG. In this regime we do not have a matching converse result. 

 \begin{corollary}\label{cor2}
Assume that 
 \begin{equation}
 \left(\muR \ge \mrstwow \quad \text{and} \quad \muT < \mtbonew \right)\hspace{1cm} \text{or} \hspace{1cm} 
 \left(\muT \ge \mtsonew \quad \text{and} \quad \muR < \mrbtwow\right).
 \end{equation}
For any $\S^{(F)} \in[ 0, \alpha \cdot \frac{\L}{2}]$, where  $\alpha$ is defined in \eqref{eq:alpha}, the pair $(\S^{(F)} ,\S^{(S)} )$ lies in the optimal MG region $\mathcal{S}^\star(\muT,\muR,\D)$ if, and only if, it is  in  the trapezoid  described on the LHS of \eqref{eq:trapezoidal2}.
 \end{corollary}
  \begin{IEEEproof}
 	Follows directly by Theorem~\ref{lemma1}, see \eqref{eq:trapezoidal2}, and  by Proposition~\ref{conv:sym}, and because the sum $\S^{(F)}_{\text{sym},3}(\alpha)+ \S^{(S)}_{\text{sym},3}(\alpha)=\L \cdot \frac{\D+1}{\D+2}$ coincides with the maximum sum MG.
 \end{IEEEproof}

 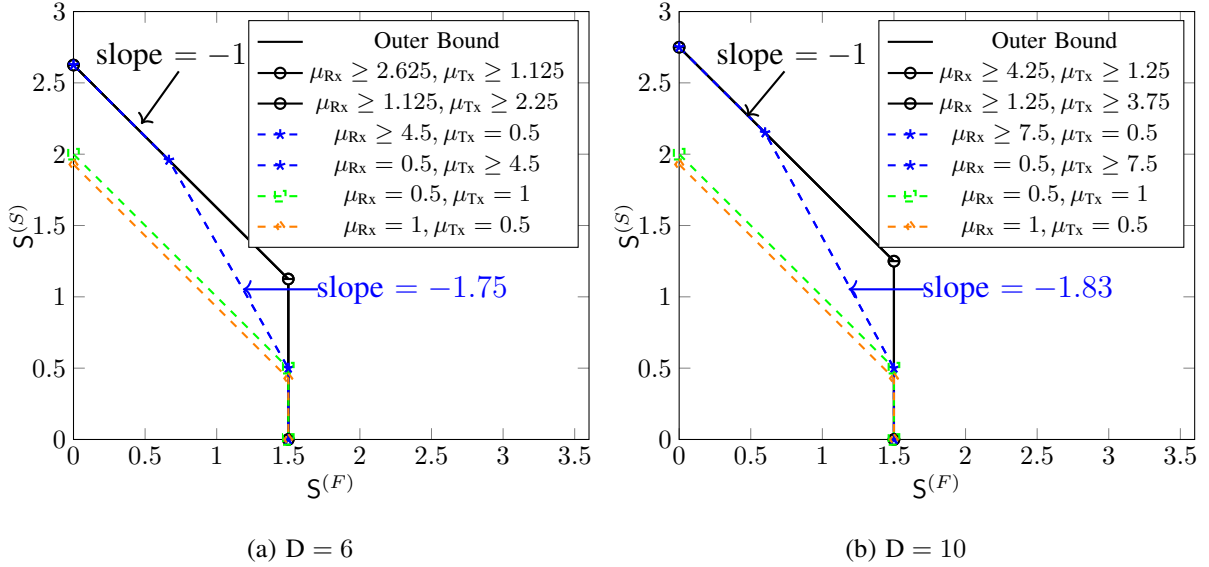
\begin{figure}[t]
 	\centering
 	\hspace{-0.2cm}
 	\begin{subfigure}{0.5\textwidth}	
 	\centering
 	\begin{tikzpicture}[scale=1]
 	\begin{axis}[
 	xlabel={\small {$\S^{(F)}$ }},
 	ylabel={\small {$\S^{(S)}$ }},
 	xlabel style={yshift=.5em},
 	ylabel style={yshift=-1.25em},
 	xmin=0, xmax=3.6,
 	ymin=0, ymax=3,
 	xtick={0,0.5,1,1.5,2,2.5,3, 3.5},
 	ytick={0,0.5,1,1.5,2,2.5,3},
 	yticklabel style = {font=\small,xshift=0.25ex},
 	xticklabel style = {font=\small,yshift=0.25ex},
 	]
 	\addplot[color=black,thick]coordinates {(0,2.625)(1.5,1.125) (1.5,0)};	
 	\addplot[color=black,mark=halfcircle,thick]coordinates {(0,2.625)(1.5,1.125) (1.5,0)};
 	\addplot[color=black,mark=halfcircle,thick]coordinates {(0,2.625)(1.5,1.125) (1.5,0)};
 	\addplot[color=blue,mark=star,thick, dashed]coordinates{(0,2.625)(0.6666,1.9583)(1.5,0.5) (1.5,0)};
 	\addplot[color=blue,mark=star,thick, dashed]coordinates{(0,2.625)(0.6666,1.9583)(1.5,0.5) (1.5,0)};

	\addplot[color=green,mark=square, thick,dashed]coordinates{(0,2.0)(1.5,0.5) (1.5,0)};
\addplot[color=orange,mark=diamond,thick, dashed]coordinates{(0, 1.928 )(1.5,0.428) (1.5,0)};

 	\legend{{\footnotesize Outer Bound}, {\footnotesize $ \muR\ge 2.625, \muT\ge 1.125$},{\footnotesize $ \muR\ge 1.125, \muT\ge 2.25$},{\footnotesize $\muR \ge 4.5, \muT =0.5$},  {\footnotesize $\muR = 0.5, \muT \ge 4.5$},{\footnotesize $\muR = 0.5, \muT =1$},{\footnotesize $\muR = 1, \muT =0.5$}}  
 	
 	\end{axis}
 	
 	\node[draw =none] (s2) at (1.3,5.1) { slope $=-1$};
 	\draw [->,  thick ] (1.4,4.9)--(0.9,4.2);
 \node[draw =none, blue] (s2) at (4.5,2 ) { slope $=-1.75$};
 	\draw [->, blue, thick ] (3.25,2)--(2.25,2);
 	\end{tikzpicture}
 	\caption{$\D = 6$}
 	\label{fig5a}
 	\end{subfigure}
 	\hspace{-0.5cm}
 	\begin{subfigure}{0.5\textwidth}
 	\centering
 	\begin{tikzpicture}[scale=1]	
 	\begin{axis}[
 	xlabel={\small {$\S^{(F)}$ }},
 	ylabel={\small {$\S^{(S)}$ }},
 	xlabel style={yshift=.5em},
 	ylabel style={yshift=-1.25em},
 	xmin=0, xmax=3.6,
 	ymin=0, ymax=3,
 	xtick={0,0.5,1,1.5,2,2.5,3, 3.5},
 	ytick={0,0.5,1,1.5,2,2.5,3},
 	yticklabel style = {font=\small,xshift=0.25ex},
 	xticklabel style = {font=\small,yshift=0.25ex},
 	]
 	\addplot[color=black,thick]coordinates {(0,2.7500)(1.5,1.2500) (1.5,0)};	
 	\addplot[color=black,mark=halfcircle,thick]coordinates {(0,2.7500)(1.5,1.2500) (1.5,0)};
 	\addplot[color=black,mark=halfcircle,thick]coordinates {(0,2.7500)(1.5,1.2500) (1.5,0)};
 	\addplot[color=blue,mark=star,thick, dashed]coordinates{(0,2.75)(0.6,2.15)(1.5,0.5) (1.5,0)};
 	\addplot[color=blue,mark=star,thick, dashed]coordinates{(0,2.75)(0.6,2.15)(1.5,0.5) (1.5,0)};

%
%
 
 	 	\addplot[color=green,mark=square, thick,dashed]coordinates{(0,2.0)(1.5,0.5) (1.5,0)};
 	 	\addplot[color=orange,mark=diamond,thick, dashed]coordinates{(0, 1.928 )(1.5,0.428) (1.5,0)};

 	\legend{{\footnotesize Outer Bound}, { \footnotesize $ \muR\ge 4.25, \muT\ge 1.25$},{\footnotesize $ \muR\ge 1.25, \muT\ge 3.75$},{\footnotesize $\muR \ge7.5, \muT =0.5$},  {\footnotesize $\muR = 0.5, \muT \ge 7.5$},{\footnotesize $\muR = 0.5, \muT =1$},{\footnotesize $\muR = 1, \muT =0.5$}}  
 	
 	\end{axis}
 	
 	\node[draw =none, blue] (s2) at (4.5,2 ) { slope $=-1.83$};
 	\draw [->, blue, thick ] (3.25,2)--(2.25,2);
 	
 	\node[draw =none] (s2) at (1.5,5.1) { slope $=-1$};
 	\draw [->,  thick ] (1.4,4.9)--(0.9,4.3);
 	
 	\end{tikzpicture}
 	\caption {$\D = 10$}
 	\label{fig5b}
 	\end{subfigure}
 	
 	\caption{Inner and outer bounds on $\mathcal{S}^\star (\muT, \muR, \D)$ for the symmetric Wyner network for  different values of $\muR$ and $\muT$, and for $\L = 3$, a) $\D=6$,  and b) $\D = 10$.}
 	\label{fig10}
 	\vspace{-0.5cm}
 \end{figure}

Figure~\ref{fig10} illustrates the inner and outer bounds (Theorem~\ref{lemma1} and Proposition~\ref{conv:sym}) on the MG region with $\D=6$ and $\D = 10$, and different values of  $\muR$ and $\muT$. As can be seen in Figure~\ref{fig5a} and as also explained in Corollary~\ref{cor1}, when $\muR\ge 2.625$ and $\muT \ge 1.125$, or when $\muR \ge 1.125$ and $\muT\ge 2.25$ the inner bound in \eqref{eq:trapezoidal1} and the outer bound match. In the former case, the inner bound is achievable using the scheme in Subsection~\ref{sub:symbothR} based on   CoMP reception, and in the latter case it is achievable using the  scheme in Subsection~\ref{sub:symbothT} based on CoMP transmission.  As explained in Corollary~\ref{cor2}, when  only one of the two cooperation prelogs is large and the other small  (e.g., $\muR \ge 4.5$ and $\muT =0.5$; or $\muT \ge 4.5$ and $\muR = 0.5$) the inner bound in \eqref{eq:trapezoidal2} matches the outer bound of Proposition~\ref{conv:sym} only for $\S^{(F)} < \alpha \cdot \frac{\L}{2}$,  where $\alpha$ is defined in \eqref{eq:alpha}. For larger values of $\S^{(F)}$,  the maximum ``slow" MG $\S^{(S)}$ achieved by our schemes  decreases linearly  with $\S^{(F)}$. For example, for $\D = 6$ and  $(\muR \ge 4.5, \muT =0.5)$ or $(\muT \ge 4.5,\muR = 0.5)$ when $\S^{(F)} \geq \alpha \frac{\L}{2}$ increases  by $\Delta$ then $\S^{(S)}$ decreases by  approximately $1.75\Delta$ and  the sum-MG      by $0.75 \Delta$. The behaviour changes again when both $\muR$ and $\muT$ are moderate or small, e.g., $\muR = 0.5$ and $\muT = 1$ or $\muR = 1$ and $\muT = 0.5$. In this case, the sum-MG achieved by our inner bound is  constant over all  regimes of $\S^{(F)}$.  We finally notice that in these small cooperation-prelog regimes our inner bounds remain unchanged for $\D=6, 8,10$. The reason is that in this regime, even when $\D>6$, it is more advantageous to reduce the number of cooperation rounds  to $6$ in order to satisfy the cooperation prelogs  than to time-share different schemes with $\D>6$ cooperation rounds. 


\section{Hexagonal Network} \label{sec:hexa}
\begin{figure}[t]
  \vspace*{-2ex}
  	\centering
\begin{tikzpicture}[scale=0.5, >=stealth]
  	\centering
  	\foreach \j in {-1,0,1}
  	\foreach \i in {-1,0,1}{
  	  \draw  (2+3 +0.5 + 3*\j,1.7321+2*1.7321+0.8660+1.7321*\i) --(3*1+3*cos{60}+3*\j,2*sin{60}*3+sin{60}+1.7321*\i)--(3*1+1+3*\j,2*sin{60}*3+1.7321*\i)--+(-60:1)--(2+3 +0.5+3*\j,1.7321+1*1.7321+0.8660+1.7321*\i)--(3*1+1+2+3*\j,2*sin{60}*3+1.7321*\i) --(2+3 +0.5+3*\j,1.7321+2*1.7321+0.8660+1.7321*\i);

  	       \draw [fill=black] (2+3 + 3*\j,1.7321+2*1.7321+1.7321*\i)  circle (0.15);
  	       
  	        \foreach \a in {-30, 30, 150,-150, -90} {   
  	     \draw[dashed] (2+3 + 3*\j,1.7321+2*1.7321+1.7321*\i)--+(\a:1.7321);
  	     }
  	     
  	      \foreach \a in {90} {   
  	     \draw[dashed] (2+3 + 3*\j,1.7321+2*1.7321+1.7321*\i)--+(\a:1.7321*0.5);
  	     }
}

\foreach \j in {-1,0,1}
  	\foreach \i in {-1,0,1}{ 	       
	       \draw  (2+3 +0.5 + 1.5+3*\j,1.7321+1*1.7321+0.8660+ 0.5*1.7321+1.7321*\i) --(3*1+3*cos{60}+ 1.5+3*\j,2*sin{60}*3+sin{60}-1.7321+ 0.5*1.7321+1.7321*\i)--(3*1+1+ 1.5+3*\j,2*sin{60}*3-1.7321+ 0.5*1.7321+1.7321*\i)--+(-60:1)--(2+3 +0.5+ 1.5+3*\j,1.7321+1*1.7321+0.8660-1.7321+ 0.5*1.7321+1.7321*\i)--(3*1+1+2+ 1.5+3*\j,2*sin{60}*3-1.7321+ 0.5*1.7321+1.7321*\i) --(2+3 +0.5+ 1.5+3*\j,1.7321+2*1.7321+0.8660-1.7321+ 0.5*1.7321+1.7321*\i);
	     
 \foreach \a in {-30, 30, 90, 150,-150} {
	        \draw[dashed] (2+3 + 1.5+3*\j,1.7321*\i+2*1.7321+0.5*1.7321)--+(\a:1.7321);
	        }
	         \foreach \a in {-90} {
	        \draw[dashed] (2+3 + 1.5+3*\j,1.7321*\i+2*1.7321+0.5*1.7321)--+(\a:1.7321*0.5);
	        }
	          \draw [fill= black] (2+3 + 1.5+3*\j,1.7321*\i+2*1.7321+0.5*1.7321)  circle (0.15);
	       }

	   \draw[red, very thick, ->] (2+3 + 1.5+3*0,1.7321*0+2*1.7321+0.5*1.7321)--+(90:1.7321);
	             \draw[red, very thick, ->] (2+3 + 1.5+3*0,1.7321*0+2*1.7321+0.5*1.7321)--+(-30:1.7321);
\node[draw =none] at (2+3 + 1.5+3*0 + 0.4,1.7321*0+2*1.7321+0.5*1.7321-0.45-0.1) {\large ${\color{red}\mathbf{e}_x}$};
\node[draw =none] at (2+3 + 1.5+3*0 + 0.35+0.1,1.7321*0+2*1.7321+0.5*1.7321+0.5) {\large ${\color{red}\mathbf{e}_y}$};      

  	\end{tikzpicture} 
	 \vspace*{-1ex}
  	\caption{Illustration of the hexagonal   network. Small circles indicate Txs and Rxs,  black solid lines  the cell borders, and  black dashed lines interference between cells. }
  	\label{fig1-1t}
  	 \vspace*{-2ex}
  \end{figure}
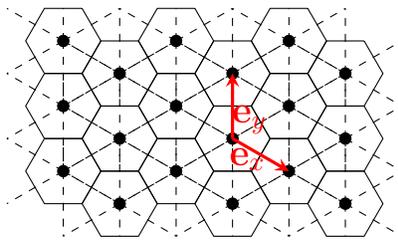

Consider a  network with $K$ hexagonal cells, where each cell consists of one single mobile user (MU) and one BS. The signals of users that lie in a given cell interfere with the signals sent in the $6$ adjacent cells. 
The interference pattern of our network  is depicted by the black dashed lines in Fig.~\ref{fig1-1t}, i.e., 
 the interference set $\mathcal I_k$ contains the indices of the $6$ neighbouring cells whose signals interfere with cell $k$. The input-output relation of the network is as in \eqref{ykn}.

Each Rx~$k$ (BS of a cell) can cooperate with the six Rxs in the adjacent cells, i.e., $|\Nrk| = 6$. Thus, the number of Rx-cooperation links $\q_{K,\Rx}$ in this network is approximately equal to $6K$ (up to edge effects). Similarly, each Tx (MU of a cell) can cooperate with the six Txs in the adjacent cells and thus $|\Ntk| = 6$ and $\q_{K,\Tx} \approx 6K$. 

To describe the setup and our schemes in  detail, we parametrize the locations of the Tx/Rx pair in the $k$-th cell by a  number $o_k$ in the complex plane $\mathbb{C}$. Introducing the coordinate vectors 
 \begin{equation}
 \mathbf{e}_x = \frac{\sqrt{3}}{2}- \frac{1}{2}i \quad \text{and} \quad  \mathbf{e}_y = i,
 \end{equation}
as in Figure~\ref{fig1-1t}, the position $o_k$ of Tx/Rx pair $k$ can be associated with  integers   $(a_k,b_k)$ satisfying
  \begin{equation}
  o_k \triangleq a_k\cdot  \mathbf{e}_x + b_k \cdot  \mathbf{e}_y.
  \end{equation}
The interference set $\mathcal I_k$ and the neighbouring sets can then be expressed as
\begin{IEEEeqnarray}{rCl}\label{eq:neighbour_Hex}
\Ntk=\Nrk=\mathcal I_k&=&\big\{ k' \colon \quad  |a_k-a_{k'}|=1 \quad \textnormal{and} \quad |b_k-b_{k'}|=1 \nonumber \\
&& \hspace{4.2cm}  \textnormal{and} \quad |a_k-a_{k'}-b_k + b_{k'}|= 1\big\}. 
\end{IEEEeqnarray}

For simplicity we assume an even-valued $\D$ satisfying
 \begin{equation} \label{eq:59}
\maw{\frac{\D}{2} -1 \mod 3 = 0.}
 \end{equation}
Other cases can be treated in a similar way.  

We  specify the Tx/Rx set associations for the   schemes in Section~\ref{coding}. See  Appendix~\ref{app-A} for a detailed analysis. For  the no-cooperation scheme in Subsection~\ref{sub:onlyfast} choose 
\begin{equation}\label{eq:Tactive}
\mathcal{T}_{\textnormal{active}} =\{ k \in [K]\colon \quad (a_k+b_k) \mod 3=0\}
\end{equation}
and   $\mathcal{T}_{\textnormal{silent}}=[K]\backslash \mathcal{T}_{\textnormal{active}}$. The corresponding cell association is shown in Figure~\ref{fig7-tta}  where active cells are  in yellow and silenced  in white. By 
\eqref{sfm}, the sum-MG achieved by this scheme is
\begin{equation}\label{hexa-fast} \S_{\text{no-coop}}=\frac{\L}{3}.
\end{equation}

We next explain the Tx/Rx set association for sending only ``slow" messages as in Subsection~\ref{sub:onlyslow}, see also \cite{samet}. Set $\tau =\frac{\D}{2} +1$ and  choose Tx~$k$ (Rx~$k$) as a master Tx (Rx), if it belongs to 
\begin{IEEEeqnarray}{rCl}\label{eq:master}
\mathcal{T}_{\textnormal{master}} &= &\left\{ k \in [K]\colon 
( a_k \mod \tau=0) \; \textnormal{and}\; ( b_k \mod \tau=0)\; \textnormal{and} \; ( |a_k+b_k| \mod 3 \tau=0) \right\}. \IEEEeqnarraynumspace
\end{IEEEeqnarray}
To describe the silenced $\mathcal{T}_{\textnormal{silent}}$, we define for any integers $x$ and $\tau\geq 0$:
\begin{equation}\label{eq:tau_shift}
x_{[-\tau,2\tau]} \triangleq ( (x +\tau) \mod 3 \tau ) -\tau,
\end{equation} 
where $\mod$ denotes the standard modulo operator. In fact, the operator $x_{[-\tau,2\tau]}$ ressembles the standard $\mod 3 \tau$ operator, but it shifts every number into the interval $[-\tau, 2 \tau)$ and not into $[0,3\tau)$. We then set
\begin{equation}\label{eq:Tsilent}
\mathcal{T}_{\textnormal{silent}}=\left\{k \colon \max \{|a_{k[-\tau, 2 \tau)}|,|b_{k[-\tau, 2 \tau)}|, |a_{k[-\tau, 2 \tau)}-b_{k[-\tau, 2 \tau)}|\} =\tau   \right\}
\end{equation}
and $\mathcal{T}_{\textnormal{slow}}= [K]\backslash \mathcal{T}_{\textnormal{silent}}$. 
 Figure~\ref{fig7-ttb}  shows the proposed cell association for $\D=6$: blue or yellow are the active ``slow" cells and  white the silenced cells. Master Txs (Rxs) are depicted with green borders. We observe that the choice in \eqref{eq:Tsilent} silences all Tx/Rx pairs  which lie $\frac{\D}{2}+1$  hops away from a master Tx/Rx pair. As we show in  Appendix~\ref{app-A},
  by \eqref{mgs} and \eqref{ssm}, this choice establishes an achievable MG pair of $(\S^{(F)} = 0, \S^{(S)}  = \ssmh)$, where  
\begin{equation} \label{slow-hexa}
\ssmh \triangleq \L \cdot \frac{4 + 3\D(\D+2)}{3(\D+2)^2}.
\end{equation} 
Moreover, by \eqref{mrstwo} and \eqref{mtsone}, with CoMP reception or CoMP transmission the scheme requires  average Rx- or Tx-cooperation prelogs  equal to 
\begin{equation}
\mtsoneh =\mrstwoh = \L\cdot \frac{\D(\D+1)}{9(\D+2)}.
\end{equation}
 \input{Cell_association_hexa2.tex}
Finally, we turn to the scheme that sends both ``fast" and ``slow" messages in Subsections~\ref{sub:both} and \ref{sub:both2}. Here, we set  $\tau =\frac{\D}{2}$ and choose the set of master Txs (Rxs) as in  \eqref{eq:master}, but for this new value of $\tau$. Similarly,  we choose the silenced set $\mathcal{T}_{\textnormal{silent}}$ as in  \eqref{eq:Tsilent} but again for  the new value $\tau=\frac{\D}{2}$. The ``fast" transmit set $\mathcal{T}_{\textnormal{fast}}$ is chosen in the same way as $\mathcal{T}_{\textnormal{active}}$ in \eqref{eq:Tactive}, and  $\mathcal{T}_{\textnormal{slow}}=\K \backslash \{\mathcal{T}_{\textnormal{silent}} \cup \mathcal{T}_{\textnormal{fast}}\}$. The cell association is depicted in Figure~\ref{fig7-ttb} for $\D=8$, where ``fast" cells  are in yellow, ``slow" cells in blue, and master cells are designated with green borders. As show in  Appendix~\ref{app-A}, 
by \eqref{eq:MG}, the proposed cell association 
achieves the MG pair $(\S^{(F)} = \sfbh, \S^{(S)}=\ssbh)$ where
\begin{equation} \label{both-hexa}
\sfbh \triangleq  \frac{\L}{3} \left ( 1 - \frac{2(\D-2)}{\D^2} \right ) \quad \text{and} \quad \ssbh \triangleq \frac{2\L}{3} \left (1 - \frac{2}{\D} \right ), 
\end{equation}
 and by \eqref{mtbone} and  \eqref{mrbone} the  average Tx- and Rx-cooperation prelogs  with CoMP reception are
\begin{IEEEeqnarray}{rCl}\label{eq:mutmurhexa1}
 \mtboneh  \triangleq   \L \cdot \frac{(\D-2)(3\D-4)}{9\D^2} \quad \text{and} \quad  \mrboneh  \triangleq  \L \cdot \frac{2\D^3 + 3\D^2 - 30 \D +32}{27\D^2},
\end{IEEEeqnarray}
and  with CoMP transmission they are 
\begin{IEEEeqnarray}{rCl}\label{eq:mutmurhexa2}
 \mtbtwoh  \triangleq  \L \cdot \frac{2\D^3 -12\D -28}{27\D^2} \quad \text{and} \quad \mrbtwoh \triangleq  \L \cdot \frac{(\D-2)(3\D-4)}{9\D^2}.
\end{IEEEeqnarray}


\subsection{Achievable MG Region}
Recall the definitions of $\sfmh$, $\ssmh$, $\sfbh$, $\ssbh$ in  \eqref{hexa-fast}, \eqref{slow-hexa}, and  \eqref{both-hexa}, and the definitions of $ \mtboneh$, $\mrboneh$, $ \mtbtwoh$ and $\mrbtwoh$ in \eqref{eq:mutmurhexa1} and \eqref{eq:mutmurhexa2}. Define
\begin{IEEEeqnarray} {rCl}\label{eq:alpha2h}
\alpha_1 \triangleq \max\left\{   \min\left\{ \frac{\muT}{\mtboneh}, \frac{\muR}{\mrboneh}\right\},\;  \min\left\{ \frac{\muT}{\mtbtwoh}, \frac{\muR}{\mrbtwoh}\right\} \right\}, \quad
\alpha_2 \triangleq \max\left\{    \frac{\muT}{\mtsoneh}, \frac{\muR}{\mrstwoh}\right\}. \IEEEeqnarraynumspace
\end{IEEEeqnarray}
Also, define 
\begin{IEEEeqnarray}{rClrCl}
 \S^{(F)}_{\text{hexa},1} (\alpha_1) &\triangleq& \alpha_1 \sfbh, \qquad  \qquad & \S^{(S)}_{\text{hexa},1}(\alpha_1) &\triangleq & \alpha_1 \ssbh + (1- \alpha_1) \ssmh,\\
  \S^{(F)}_{\text{hexa},2}(\alpha_1) &\triangleq& \alpha_1 \sfbh + (1-\alpha_1)\sfmh, \qquad  \qquad &  \S^{(S)}_{\text{hexa},2}(\alpha_1) & \triangleq & \alpha_1 \ssbh, \\
   \S^{(S)}_{\text{hexa}}(\alpha_2) &\triangleq& \alpha_2 \ssmh + (1 - \alpha_2)\ssnch.
\end{IEEEeqnarray}

 \begin{theorem} [Achievable MG Region: Hexagonal Model] \label{lemma2} 
Assume $\D \ge 2$\maw{, even, and $\frac{\D}{2} -1 \mod 3 = 0.$}

 		 When $\muR \geq \max \{\mrboneh, \mrstwoh \}$ and $\muT \ge \mtboneh$; or when $\muT \geq \max \{\mtbtwoh, \mtsoneh \}$ and $\muR \ge \mrbtwoh$;  \maw{then:}
 		\begin{IEEEeqnarray}{rCl}\label{eq:quad1}
 		\textnormal{convex hull}\Big(  (0,0), \ (0, \ssmh), \  (\sfbh, \ssbh),  \  (\sfmh, 0) \Big) \subseteq \mathcal{S}^\star(\muT,\muR,\D). \IEEEeqnarraynumspace
 		\end{IEEEeqnarray}

 When $\mrboneh \le \muR < \mrstwoh$ and $\muT \ge \mtboneh$; or when $\mtbtwoh \le \muT < \mtsoneh$ and $\muR \ge \mrbtwoh$;  \maw{then:}
\begin{IEEEeqnarray}{rCl}\label{eq:quad1}
 		\textnormal{convex hull}\Big(  (0,0), \ (0, \sfbh + \ssbh ), \  (\sfbh, \ssbh),  \  (\sfmh, 0) \Big)\subseteq \mathcal{S}^\star(\muT,\muR,\D). \IEEEeqnarraynumspace
		\end{IEEEeqnarray}

When  $\muR \ge \mrstwoh$ and $\muT < \mtboneh$; or when  $\muT \ge \mtsoneh$ and $\muR < \mrbtwoh$;   \maw{then:}
 		\begin{IEEEeqnarray}{rCl}\label{eq:quad1}
 		\textnormal{convex hull}\Big(  (0,0), \ (0, \ssmh), &&\  (\S^{(F)}_{\text{hexa},1} (\alpha_1), \S^{(S)}_{\text{hexa},1} (\alpha_1)), \nonumber \\
 		&& \ (\S^{(F)}_{\text{hexa},2} (\alpha_1), \S^{(S)}_{\text{hexa},2} (\alpha_1)), \  (\sfmh, 0) \Big)\subseteq \mathcal{S}^\star(\muT,\muR,\D). \IEEEeqnarraynumspace
 		\end{IEEEeqnarray}

 When $\muR <\mrboneh $  or {\maw{when} $\muT < \mtbtwoh$,  then:}
 		\begin{IEEEeqnarray}{rCl}\label{eq:quad1}
 		\textnormal{convex hull}\Big(  (0,0), \ (0, \S^{(S)}_{\text{hexa}} (\alpha_2)), \  (\S^{(F)}_{\text{hexa},2} (\alpha_1), \S^{(S)}_{\text{hexa},2} (\alpha_1)),  \  (\sfmh, 0) \Big) \subseteq \mathcal{S}^\star(\muT,\muR,\D). \IEEEeqnarraynumspace
 		\end{IEEEeqnarray}


 \end{theorem}
 \begin{figure}[t]
 	\centering
 	\begin{tikzpicture}[scale=0.95]
 	
 	\begin{axis}[
 	xlabel={\small {$\S^{(F)}$ }},
 	ylabel={\small {$\S^{(S)}$ }},
 	xlabel style={yshift=.5em},
 	ylabel style={yshift=-1.25em},
 	xmin=0, xmax=1.7,
 	ymin=0, ymax=2.5,
 	xtick={0,0.5,1,1.5,2},
 	ytick={0,0.5,1,1.5,2},
 	yticklabel style = {font=\small,xshift=0.25ex},
 	xticklabel style = {font=\small,yshift=0.25ex},
	legend style={at={(1.4,.7)}},
 	]
 		
 	\addplot[color=black,mark=halfcircle,thick]coordinates {(0,2.4400)(0.8125,1.5) (1,0)};
 	\addplot[color=black,mark=halfcircle,thick]coordinates {(0,2.4400)(0.8125,1.5) (1,0)};
 	
 	\addplot[color=blue,mark=star,thick, dashed]coordinates {(0,2.3125)(0.8125,1.5) (1,0)};
 	\addplot[color=blue,mark=star,thick, dashed]coordinates {(0,2.3125)(0.8125,1.5) (1,0)};
 	
	\addplot[color=green,mark=square, thick,dashed]coordinates{(0,2.4400)(0.13,2.2896)(0.9700, 0.24) (1,0)};
	\addplot[color=green,mark=square, thick,dashed]coordinates{(0,2.4400)(0.13,2.2896)(0.9700, 0.24) (1,0)};

%
	\addplot[color=orange,mark=diamond,thick, dashed]coordinates{(0,1.6)(0.88,0.96) (1,0)};
	\addplot[color=red, thick, dashed]coordinates{(0,1.6)(0.8928,0.8571) (1,0)};

\small {
 	\legend{{\small $ \muR\ge2.4, \muT\ge0.6$},{ $ \muR\ge0.63, \muT\ge2.4$},  { $1.7\le\muR <2.4 , \muT \ge 0.6$},{ $\muR \ge 0.6, 1.5 \le \muT < 2.4 $}, {$\muR\ge 2.4, \muT =0.1$},{$\muR = 0.1, \muT \ge 2.4$},{$\muR = 0.5, \muT =1$},{$\muR = 1, \muT =0.5$}} } 
 	
 	\end{axis}

 	 \node[draw =none] (s2) at (1.9,5.2 ) { \small slope $=-1.15$};
 	 \draw [->, thick] (1.85,5.1)--(1.85,4.3);
 	\end{tikzpicture}

 	\caption{Inner bounds on $\mathcal{S}^\star (\muT, \muR, \D)$ for the hexagonal model  for $\D = 8$,  $\L = 3$ and different values of $\muR$ and $\muT$.}
 	\label{fig5.6}
 	\vspace{-0.5cm}
 \end{figure}
 
\mw{The maximum achievable ``fast" MG is $\frac{\L}{3}$. As previously mentioned, in this paper we do not consider the impractical ergodic interference alignment with infinite symbol extensions, which achieves a ``fast" MG of $\frac{\L}{2}$.}
Figure~\ref{fig5.6} illustrates the inner  bounds (Theorem~\ref{lemma2}) on the MG region for $\D=8$, and different values of  $\muR$ and $\muT$. We observe that, unlike  for Wyner's symmetric model, the sum-MG of this network always decreases as $\S^{(F)}$ increases, irrespective of the cooperation prelogs $\muT,\muR$. 
\maw{Moreover,   maximum $\S^{(F)}$ is only achieved for    $\S^{(S)}=0$.}  

In  Figure~\ref{fig5.6}, we can distinguish 4 behaviours for the achieved MG region:  1) If both $\muR$ and $\muT$ are above given thresholds, for $\D=8$ and either $(\muT \geq 0.6, \muR\geq 2.4)$ or $(\muT \geq 0.63, \muR\geq 2.4)$, then the points $(0, \S^{(S)}_{\max})$ and $(\S_{\text{both}}^{(F)}, \S_{\text{both}}^{(F)})$ are both achievable. 2) When one of the two cooperation prelogs remains very high ($\muR$ or $\muT$ larger than $2.4$) but the other one becomes relatively small, only $(0, \S^{(S)}_{\max})$ is achievable, but not $(\S_{\text{both}}^{(F)}, \S_{\text{both}}^{(F)})$.  The largest achievable $\S^{(S)}$ is thus not reduced as long as $\S^{(F)}$ remains small; for larger values of $\S^{(F)}$ the maximum achievable $\S^{(S)}$ however suffers significantly. The reason is that  our schemes that send both  ``fast" and ``slow" messages inherently require both Tx- \emph{and} Rx-cooperation of sufficiently high cooperation prelogs. 
 As a consequence, the maximum  $\S^{(S)}$ that our schemes achieve for large $\S^{(F)}$ highly depends on the smaller of the two cooperation prelogs $\muT$ and $\muR$. 
 3) When both $\muT,\muR$ are moderate,  we can still achieve the MG pair  $(\S_{\text{both}}^{(F)}, \S_{\text{both}}^{(F)})$ but not $(0, \S^{(S)}_{\max})$. In the regime of small $\S^{(F)}$ there is thus a penalty in $\S^{(S)}$ and sum MG compared to the case of high cooperation prelogs but not in the regime of large $\S^{(F)}$. 4) Finally, when both cooperation prelogs become small then neither of the two points  $(0, \S^{(S)}_{\max})$ and $(\S_{\text{both}}^{(F)}, \S_{\text{both}}^{(F)})$  is achievable anymore.

\input{sect-cell-D8-new.tex}

\vspace{-5mm}

 \section{Sectorized Hexagonal Model} \label{sec:sec}
\maw{Reconsider the cellular network with  $K$ hexagonal cells and cell coordinate system spanned by  the vectors $\mathbf{e}_x$ and $\mathbf{e}_y$   introduced in the previous section. Here, each cell consists of three sectors denoted by ``S", ``W", and ``E", see Figure~\ref{fig7}, and we also number the sectors from $1$ to $3K$. A  single $3\L$-antenna Rx (BS) is associated to each \emph{cell} and a single $\L$-antenna Tx to each \emph{sector}. Each Rx decodes the 3 ``slow" and the 3 ``fast" messages of the Txs in the $3$ sectors corresponding to its cell.  \maw{Rxs} are equipped with  directional antennas,  where each set of $\L$ antennas at a given Rx (BS)  points to one of the three sectors of its cell.  Therefore, communications from different sectors in the same cell do not interfere, see 
Fig.~\ref{fig7} where interference is depicted by dashed lines. Interference is short-range, and transmission in the grey-shaded sector of Fig.~\ref{fig7} is, e.g.,  interfered  by the transmissions in the four adjacent pink-shaded sectors. The interference set $\mathcal{I}_{\Tx, {k}'}$ of sector ${k}'$ is thus the set of indices of the $4$ adjacent sectors that lie in a different cell.

For the purpose of this section, we thus modify the setup in Section~\ref{setup} in that we have $3K$ Txs and $K$ Rxs and each Rx $k$ observes the output signals 
 $\mathbf Y_{k}^n:=(\mathbf Y_{k_1}^n ,\mathbf Y_{k_2}^n ,\mathbf Y_{k_3}^n )$, where  $k_1,k_2,k_3$ denote the three sectors in cell $k$, and
\begin{equation}\label{ykn}
\mathbf Y_{k_i}^n =  \mathsf H_{k_i,k_i} \vect X_{k_i}^n + \sum_{\hat k \in \mathcal I_{k_i}} \mathsf{H}_{\hat k ,k_i}  \vect X_{\hat k}^n + \vect Z_{k_i}^n,\qquad  i \in \{1,2,3\}.
\end{equation}
We consider \emph{per-sector MGs}, and accordingly the average rates in \eqref{eq:average_rates} are normalized with respect to $3K$ and not $K$. All other definitions of Section~\ref{setup} remain unchanged.

Each Rx~$k$ (BS of a cell) can cooperate with the  Rxs in the six adjacent cells, i.e., $|\Nrk| = 6$ and $\q_{K,\Rx} \approx 6K$. Each Tx (MU of a cell) can cooperate with the four Txs in the adjacent sectors of different cells, i.e. $|\Ntk| = 4$ and since there are $3K$ Txs, $\q_{K,\Tx} \approx 12K$. Assume  $\D$  even.

The  coding schemes and results in Section~\ref{coding} apply also to this modified setup, if $\mathcal{T}_{\text{silent}}, \mathcal{T}_{\text{active}}, \mathcal{T}_{\text{fast}}, \mathcal{T}_{\text{slow}}\subseteq [3K]$ and the MG results \eqref{eq:MG}, \eqref{ssm}, and \eqref{sfm} are normalized with respect to $3K$ and not $K$.
We only consider CoMP reception, and thus $\mathcal{T}_{\textnormal{master}}\subseteq  [K]$.}

\subsection{\maw{Tx/Rx Set Associations and MG Region}} \label{sub:sectset}

We specify the Tx/Rx set associations for our \mw{schemes of Section~\ref{coding}}. For  the no cooperation scheme,  define the active set $\mathcal{T}_{\textnormal{active}}$ as the set of either the ``W" sectors, the ``E" sectors, or the ``S" sectors of all cells. 
This achieves the  sum-MG 
\begin{equation} \label{sec-fast}
\S_{\textnormal{no-coop}} \triangleq \frac{\L}{3}.
\end{equation} 

For the cooperative schemes,  we pick the  set of \textit{master cells} $\mathcal{T}_{\textnormal{master}}$ as in \eqref{eq:master} for  $\tau = \frac{\D}{2}$. 
 Unlike in the hexagonal model in Section \ref{sec:hexa}, it suffices to silence certain sectors of layer  $\D/2$ around each master cell. Consider the subnet that has its master cell $k_{\textnormal{master}}$ at the origin $a_{k_{\textnormal{master}}}=b_{k_{\textnormal{master}}}=0$, for which  we keep active all 3 sectors of the corner cells in layer $\D/2$ that have coordinates $(a_k=\D/2, b_k=0)$, $(a_k=0, b_k =\D/2)$, and $(a_k=-\D/2, b_k=-\D/2)$, and we silence all 3 sectors of the remaining 3 corner cells of this layers, which have coordinates $(a_k=\D/2, b_k =\D/2)$, $(a_k=-\D/2, b_k=0)$, and $(a_k=0, b_k=-\D/2)$. We further silence in this layer $\D/2$   the ``S" sector of all non-corner cells with  coordinates  $|b_k|=\D/2$ and $\textnormal{sign}(a_k)=\textnormal{sign}(b_k)$;  the ``E" sector of all non-corner cells with coordinates $|a_k|=\D/2$ and $\textnormal{sign}(a_k)=\textnormal{sign}(b_k)$; and the ``W" sector  of all non-corner cells  with  coordinates $\textnormal{sign}(a_k)\neq\textnormal{sign}(b_k)$.  As for the hexagonal model, all  Txs that lie less than $\D/2$ cell hops from a master cell are  kept active. The proposed sector association splits the entire network into equal non-interfering subnets (up to  edge effects that vanish as $K\to \infty$), each consisting of a master cell, all sectors of the cells in the $\D/2-1$ surrounding layers, and none or one sector in each cell of layer~$\D/2$.
 The proposed cell and sector association is shown in Figure~\ref{fig5.10b} for $\D=8$, where yellow and blue sectors are active and white are silenced. The borders of the subnets are shown by red lines. 
 
As shown in  Appendix~\ref{app-B}, when sending only ``slow" messages the proposed sector association achieves $(\S^{(F)} = 0, \S^{(S)}  = \ssms)$ where 
\begin{equation} \label{slow-sec}
\ssms \triangleq \L \cdot \frac{3\D-2}{3\D},
\end{equation} 
and it requires an average  Rx-cooperation prelog of
\begin{equation}\label{eq:mrstwos}
\mrstwos =  \L \cdot \frac{(\D-1)}{3}.
\end{equation}

 In the  scheme sending both ``fast" and ``slow" messages, the Txs in the ``yellow" sectors  of Figure~\ref{fig5.10b} send ``fast" messages and  the Txs in the ``blue" sectors send ``slow" messages. We describe the cell association more formally for a subnet whose master cell is at the origin. All other subnets are equal. 
 All active sectors in layer-$\D/2$ of this subnet send ``fast" messages, but all sectors in the cells satisfying one of the three following conditions only send ``slow" messages: ($a_k\geq 0$ and $b_k=0$) or ($a_k=0$ and $b_{k}\geq 0$) or ($a_k=b_k\leq 0$).
 All other cells have exactly one ``fast" sector and two ``slow" sectors. 
 Specifically, cells with $a_k, b_k >0$ send a ``fast" message in their ``W" sector; 
 cells with $a_k<0$ and $b_k>a_k$ send a ``fast" message in their ``S" sector; and
   cells with $b_k<0$ and $a_k>b_k$ send a ``fast" message in their ``E" sector.
 
 We prove  in Appendix~\ref{app-B} that  the proposed sector association achieves the  MG pair 
 \begin{equation} \label{both-sect}
\sfbsr \triangleq \frac{\L}{3}, \quad \text{and} \quad \ssbsr \triangleq \L \cdot \frac{2\D-2}{3 \D},
\end{equation}
and requires average Tx- and Rx-cooperation prelogs 
\begin{IEEEeqnarray}{rCl} \label{eq:mtbones}
 \mtbones  \triangleq   \L \cdot \frac{(\D-1)}{3\D} \quad \text{and} \quad  \mrbones  \triangleq \L \cdot \frac{2\D^2-5}{9\D}.
\end{IEEEeqnarray}

 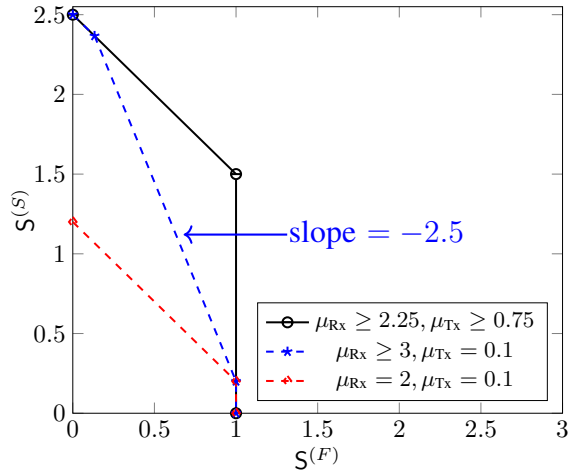
\begin{figure}[t]
 	\centering
 	\begin{tikzpicture}[scale=.95]
 	
 	\begin{axis}[
 	xlabel={\small {$\S^{(F)}$ }},
 	ylabel={\small {$\S^{(S)}$ }},
 	xlabel style={yshift=.5em},
 	ylabel style={yshift=-1.25em},
 	xmin=0, xmax=3,
 	ymin=0, ymax=2.55,
 	xtick={0,0.5,1,1.5,2,2.5,3},
 	ytick={0,0.5,1,1.5,2,2.5},
 	yticklabel style = {font=\small,xshift=0.25ex},
 	xticklabel style = {font=\small,yshift=0.25ex},
 	 legend pos=south east,
 	]
 		
 	\addplot[color=black,mark=halfcircle,thick]coordinates {(0,2.5000)(1,1.5) (1,0)};
 	
	\addplot[color=blue,mark=star, thick,dashed]coordinates{(0,2.5000)(0.1333,2.3667)(1, 0.2) (1,0)};

                      \addplot[color=red,mark=diamond,thick, dashed]coordinates {(0,1.2)(1,0.2) (1,0)};

 	\legend{{\footnotesize $ \muR\ge2.25, \muT\ge 0.75$},{\footnotesize $ \muR\ge 3, \muT=0.1$},  {\footnotesize $\muR = 2, \muT =0.1$}}  
 	
 	\end{axis}

 	 \node[draw =none, blue] (s2) at (4.25,2.5 ) { slope $=-2.5$};
 	\draw [->, blue, thick ] (3,2.5)--(1.55,2.5);
 	\end{tikzpicture}

 	\caption{Inner bounds on $\mathcal{S}^\star (\muT, \muR, \D)$ for the sectorized hexagonal model  for $\D = 4$,  $\L = 3$ and different values of $\muR$ and $\muT$.}
 	\label{fig5.11}
 	\vspace{-0.5cm}
 \end{figure}
\maw{Recall  definitions~\eqref{sec-fast}--\eqref{both-sect} and 
define} 
\begin{IEEEeqnarray}{rCl}
\alpha_1 \triangleq  \frac{\muT}{\mtbones} \quad \text{and} \quad  
\alpha_2 \triangleq   \min\left\{ \frac{\muT}{\mtbones}, \frac{\muR}{\mrbones}\right\},
\end{IEEEeqnarray}
\begin{IEEEeqnarray}{rClrCl}
\S^{(F)}_{\text{sec}}(\alpha_1) &\triangleq& \alpha_1 \sfbsr, \quad & \S^{(S)}_{\text{sec}}(\alpha_1)&\triangleq& \alpha_1 \ssbsr + (1- \alpha_1) \ssmsr,\\
 \maw{ \S^{(F)}_{\text{sec},1}(\alpha_2}) &\triangleq& \alpha_2 \sfbsr + (1-\alpha_2)\maw{\S_{\textnormal{no-coop}}}, \hspace{1.3cm} & \maw{\S^{(S)}_{\text{sec},1}}(\alpha_2)&\triangleq& \alpha_2 \ssbsr, \\
\maw{  \S^{(S)}_{\text{sec},2}(\alpha_2)} &\triangleq& \alpha_2 \ssmsr + (1 - \alpha_2)\maw{\S_{\textnormal{no-coop}}}.
\end{IEEEeqnarray}
The following theorem is proved in Appendix~\ref{app-B}. 
 \begin{theorem} [Achievable MG Region: Sectorized Hexagonal Model] \label{lemma3-ch5} 
Assume $\D \geq 2$ and even.

 	\begin{itemize}
 		\item When $\muR \geq \mrbones$ and $\muT \ge \mtbones$;  
 		\begin{IEEEeqnarray}{rCl}\label{eq:quad1}
 		\textnormal{convex hull}\Big(  (0,0), \ (0, \ssmsr), \  (\sfbsr, \ssbsr),  \  (\maw{\S_{\textnormal{no-coop}}}, 0) \Big) \subseteq \mathcal{S}^\star(\muT,\muR,\D). \IEEEeqnarraynumspace
 		\end{IEEEeqnarray}

\item When $\muR \ge \mrstwos$ and $\muT <  \mtbones$;
 		\begin{IEEEeqnarray}{rCl}\label{eq:quad1}
 		\textnormal{convex hull}\Big(  (0,0), \ (0, \ssmsr), && \  (\S^{(F)}_{\text{sec}} (\alpha_1), \S^{(S)}_{\text{sec}} (\alpha_1)), \notag \\
 		&& \ \maw{ (\S^{(F)}_{\text{sec},1} (\alpha_2), \S^{(S)}_{\text{sec},1} (\alpha_2)), } \  (\maw{\S_{\textnormal{no-coop}}}, 0) \Big) \subseteq \mathcal{S}^\star(\muT,\muR,\D). \IEEEeqnarraynumspace 
 		\end{IEEEeqnarray}

\item When $\muR < \mrbones$  and $\muT <  \mtbones$; 
 		\begin{IEEEeqnarray}{rCl}\label{eq:quad1}
 		\textnormal{convex hull}\Big(  (0,0),\maw{ \ (0, \S^{(S)}_{\text{sec},2} (\alpha_2)), \  (\S^{(F)}_{\text{sec},1} (\alpha_2), \S^{(S)}_{\text{sec},1}} (\alpha_2)),  \  (\maw{\S_{\textnormal{no-coop}}}, 0) \Big)\subseteq \mathcal{S}^\star(\muT,\muR,\D). \IEEEeqnarraynumspace 
 		\end{IEEEeqnarray}
\end{itemize}
 \end{theorem}
{Figure~\ref{fig5.11} illustrates the inner  bounds (Theorem~\ref{lemma3-ch5}) on the MG region for $\D=4$, and different values of  $\muR$ and $\muT$. As can be seen from this figure, when $\muR \ge 2.25$ and $\muT\ge 0.75$, there is no penalty in sum MG even at maximum ``fast'' MG. 
 }

 \section{Conclusions} \label{sec:conclusion}
We proposed a coding scheme for general interference networks that accommodates the transmission of both delay-sensitive and delay-tolerant messages.  We characterized the MG region of Wyner's symmetric network for certain parameters and derived inner bounds on the achievable MG region for general parameters, as well as for the sectorized and non-sectorized hexagonal model.
 The results for Wyner's symmetric model showed that it is possible to accommodate the largest possible MG for delay-sensitive messages, without penalizing the maximum sum MG of both delay-sensitive and delay-tolerant messages. Our proposed scheme suggests a similar behaviour for the sectorized hexagonal model, when one restricts to one-shot interference alignment. For the non-sectorized hexagonal model this does not seem to be the case, and our results always show a penalty in sum MG whenever the delay-sensitive MG is not zero. These results indicate that each network needs to be carefully analyzed  to determine whether a sum MG  penalty exists under mixed-delay traffics. Nevertheless, in this paper we proposed a joint coding scheme for mixed-delay traffics that significantly  improves the sum MG compared to a classical scheduling approach. 
 
 \maw{Our proposed coding schemes   suggest that in the regime of high delay-sensitive MGs, it is important to have sufficiently high cooperation prelogs both at the Tx- and the Rx-side} to attain the same sum MG as when only delay-tolerant messages are sent. Moreover, in this regime, Tx-cooperation seems to be slightly more beneficial under mixed-delay traffics than Rx-cooperation. 

An interesting line of future  research is to analyze the effect of delay-sensitive messages on 
generalized Wyner models with fading coefficients and
finite precision channel state information. Here also the notion of generalized degrees of freedom (GDoF) is of interest, see also \cite{wang_etal}.

 \appendices \label{app}
 \section{Analyses  for the Hexagonal Model} \label{app-A}
   We prove that the Tx/Rx set associations  proposed in  Section~\ref{sec:hexa} are permissible and we provide details on how to compute the  corresponding MG pairs and cooperation prelogs. 

\subsection{No Cooperation Scheme} \label{sub:hexa-fast}
Fig. \ref{fig7-tta} shows the active  Txs in yellow and the silenced Txs in white.  It is easily seen that transmissions in yellow cells do not interfere, as they pertain to non-neighbouring cells.

 More formally, we consider two different  Txs $k$ and $k'$ in the active set $\mathcal{T}_{\textnormal{active}}$ defined in \eqref{eq:Tactive}, and we prove by contradiction that Tx $k'$ cannot be in the neighbouring set $\mathcal I_k$ of Tx $k$.  Assume that $k' \in \mathcal I_k$. Then, by \eqref{eq:neighbour_Hex}, either $(a_{k'}=a_{k}+1,b_{k'}=b_k+1)$ or $(a_{k'}=a_{k}-1,b_{k'}=b_k-1)$. Each of these two cases however violates the active set condition \eqref{eq:Tactive}, which implies
 \begin{equation}
 (a_{k}+b_{k}) \mod 3 =0 \quad \textnormal{and} \quad (a_{k'}+b_{k'}) \mod 3 =0.
 \end{equation}
 We thus obtained the desired contradiction.

To see that the scheme achieves a sum-MG of $\S_{\text{no-coop}}$ we notice that in the limit as $K\to \infty$, the active set $\mathcal{T}_{\textnormal{active}}$ defined in \eqref{eq:Tactive} includes a third of all Txs simply because a third of the integer pairs $(a,b)$ satisfy $(a+b) \mod 3=0$ and because  each integer pair $(a,b)$ corresponds a Tx.

\subsection{Coding scheme to transmit only ``slow'' messages with CoMP reception or transmission} \label{sub:hexa-slow}


As mentioned in the main body, and as is easily seen in Figure~\ref{fig7-ttb}, the silenced set $\mathcal{T}_{\textnormal{silent}}$ consists of all cells that are exactly $\frac{\D}{2} + 1$ cell hops away from the next  master cell. All other cells  have a master cell that lies less than $\frac{\D}{2} + 1$ cell hops away. In other words, each master cell is surrounded by $\D/2$  layers of active cells  sending ``slow" messages, where in total these $\D/2$ layers contain $\sum_{i= 1}^{\D/2} 6 i  = \frac{3}{4}\D(\D+2)$ cells.  Such a subnet is then surrounded by a layer of  $6\cdot (\frac{\D}{2}+1) = 3\D+6$ silenced cells, each lying $\D/2+1$ cell hops away from the master cell.  Among these layer-$(\D/2+1)$ cells, $6$ of them (namely the corner cells) belong to  the silenced layer of three different master cells, and the remaining $3\D$ belong to the silenced layer of two master cells. One can therefore add  a set of $(6/3+3\D/2)$ silenced cells ($2$ corner cells and $3\D/2$ non-corner cells) to each of the subnets to  partition the entire network of $K$ cells into subsets of size
\begin{equation}\label{eq:s}
s \triangleq 1+ \frac{3}{4}\D(\D+2) + (6/3+3\D/2) = \frac{3(\D+2)^2}{4}.
\end{equation}

By these considerations, and because master cells themselves also send ``slow" messages,
\begin{equation}\label{lim:slow}
\lim_{K\to \infty} \frac{|\mathcal T_{\text{slow}}|}{K}= \frac{1+ \frac{3}{4}\D(\D+2)}{\frac{3}{4}(\D+2)}= \frac{4+ 3\D(\D+2)}{3(\D+2)^2},
\end{equation}
and as a result, by \eqref{mgs} and \eqref{ssm}, the proposed cell association achieves the MG pair $(\S^{(F)}=0, \S^{(S)}=\S^{(S)}_{\textnormal{max}})$ with $\S^{(S)}_{\textnormal{max}}$ defined in \eqref{slow-hexa}.

With CoMP reception, the scheme does not send any Tx-cooperation messages but only Rx-cooperation messages. To calculate the Rx-cooperation prelog, notice that for each Rx $k$ we have   $\gamma_{\Rx,k}=i\in\{1, \ldots,\D/2\}$ if Rx $k$ lies $i$ hops away from the next master cell. Fix a master cell $k_{\textnormal{master}} \in \mathcal{T}_{\textnormal{master}}$ and define $\mathcal{T}_{\textnormal{subnet}}$ as the set including this master cell as well as the $\D/2$ layers around it: 
\begin{equation}\label{eq:subnet}
\mathcal{T}_{\textnormal{subnet}} \triangleq \left \{k \colon  \max \{ | a_k- a_{k_{\textnormal{master}}}| ,\; |b_{k}- b_{k_{\textnormal{master}}}|, \; |a_{k}- a_{k_{\textnormal{master}}}-b_{k}+ b_{k_{\textnormal{master}} }|\} \leq \frac{\D}{2}\right\}. 
\end{equation}  Since in this subnet $\mathcal{T}_{\textnormal{subnet}}$   there are $6i$ Rxs with $\gamma_{\Rx,k}=i$, for each $i=1,\ldots, \D/2$:
\begin{equation}
2\sum_{k \in  \mathcal{T}_{\textnormal{subnet}}} \gamma_{\Rx,k} = 2 \sum_{i =1}^{\D/2} 6 i^2= \frac{\D(\D+2)(\D+1)}{2},
\end{equation}
and since by a sandwiching argument $K^{-1} \sum_{k \in  \mathcal{T}_{\textnormal{slow}}} \gamma_{\Rx,k} \to s^{-1} \sum_{k \in  \mathcal{T}_{\textnormal{subnet}}} \gamma_{\Rx,k}$ as $K\to \infty$:
\begin{IEEEeqnarray}{rCl}
\lim_{K\to \infty}  \frac{2\sum_{k \in  \mathcal{T}_{\textnormal{slow}}} \gamma_{\Rx,k} }{K} & =&
\frac{2\sum_{k \in  \mathcal{T}_{\textnormal{subnet}}} \gamma_{\Rx,k}}{s} \\
&=& \frac{ \D(\D+2)(\D+1)/2}{{\frac{3}{4}(\D+2)^2}} = \frac{2\D (\D+1)}{3(\D+2)}.
\end{IEEEeqnarray}
Finally, because  $\lim_{K\to \infty} \frac{\q_{K, \Rx}}{K} =6$,  according to \eqref{mrstwo}
 the required Rx-cooperation prelog  equals   
\begin{equation}
\mrstwoh = \L \cdot \frac{\D(\D+1)}{9(\D+2)}.
\end{equation}

With CoMP transmission,  this scheme does not require any Rx-cooperation messages and  consumes a Tx-cooperation prelog of $\mtsoneh = \mrstwoh$.  
\subsection{Coding Scheme to transmit both ``fast'' and ``slow'' messages with CoMP reception} \label{sub:hexa-bothR}

Consider the cell association described for this scheme in the main body of the paper. That means, 
\begin{IEEEeqnarray}{rCl}\label{eq:Tsilent2}
\mathcal{T}_{\textnormal{silent}}=\Bigg\{k \colon \max \{|a_{k[-\tau, 2 \tau)}|,|b_{k[-\tau, 2 \tau)}|, |a_{k[-\tau, 2 \tau)}-b_{k[-\tau, 2 \tau)}|\} =\frac{\D}{2}   \Bigg\},
\end{IEEEeqnarray}
where the operator $x\mapsto x_{[-\tau,2\tau]}$ is defined in \eqref{eq:tau_shift}. 
Similarly, the set of master cells is given by
\begin{IEEEeqnarray}{rCl}\label{eq:master2}
\lefteqn{\mathcal{T}_{\textnormal{master}} = \bigg\{ k \in [K]\colon 
\left( a_k \mod \frac{\D}{2}=0\right) \; \textnormal{and}\; \left( b_k \mod \frac{\D}{2}=0\right)\; }  \nonumber \\
&& \hspace{8cm} \textnormal{and} \; \left( |a_k+b_k| \mod  \frac{3\D}{2}=0\right) \bigg\}. \IEEEeqnarraynumspace
\end{IEEEeqnarray}
This choice decomposes the network into equal subnets of active cells, each one surrounding one of the master cells. For a given master cell $k_{\textnormal{master}}$, the subnet is given by:
 \begin{equation}\label{eq:subnet2}
\mathcal{T}_{\textnormal{subnet}} \triangleq \left \{k \colon  \max \{ | a_k- a_{k_{\textnormal{master}}}| ,\; |b_{k}- b_{k_{\textnormal{master}}}|, \; |a_{k}- a_{k_{\textnormal{master}}}-b_{k}+ b_{k_{\textnormal{master}} }|\} \leq \frac{\D}{2}-1\right\}.
\end{equation} 
Notice that the subnet is defined in a similar way as in \eqref{eq:subnet} in the preceding subsection, except that $\D$ is replaced by $\D-2$. That means, it contains the master cell $k_{\textnormal{master}}$ and the $\D/2-1$ cell layers (i.e., the cells with $\gamma_{\Rx,k}=1,\ldots, \D/2-1$) surrounding it. Therefore: 
\begin{equation}\label{eq:size_subnet}
|\mathcal{T}_{\text{subnet}}|= 1 +\sum_{i=1}^{\frac{\D}{2}-1} 6i=1+ \frac{3}{4 }\D (\D-2)
\end{equation}
Cell-layer $\D/2$ (the cells with $\gamma_{\Rx,k}=\D/2$) around each master cell is silenced. It   consists of $6$ corner cells, which are $\D/2$ cell hops away from 3 different master cells, and of $3\D-6$ non-corner cells, which lie $\D/2$ cell hops away from two different master cells. Similarly to the previous section, one can build a cell partitioning by simply associating a third of the corner cells and half of the non-corner cells of layer $\D/2$ to each master cell. Any subset of such a partition is then  of size (up to some edge effects that vanish as $K\to \infty$)  
\begin{equation}\label{eq:subnet_size2}
s \triangleq \left( 1+ \frac{3}{4 }\D (\D-2)\right) + \left(2+ \frac{3 \D -6}{2}\right)= \frac{3}{4}\D^2.
\end{equation}

Recall further that we chose $\mathcal{T}_{\text{fast}}= \{k \colon (a_k+b_k)\mod 3 =0\}$ and  $\Ts= [K] \backslash ( \Tf \cup \mathcal{T}_{\text{silent}})$. 

Since all subnets are equal (there can be some edge effects that vanish as $K\to \infty$), by some sandwiching arguments, we obtain that   $K^{-1}  {|\Tf |} \to s^{-1} |\mathcal{T}_{\text{subnet}} \cap \Tf |$  and $K^{-1} {|\Ts |} \to s^{-1} |\mathcal{T}_{\text{subnet}} \cap \Ts |$ as $K \to \infty$. 
By the following Lemma~\ref{eq:lemma_sub}, we then obtain
 the asymptotic ratios: 
\begin{IEEEeqnarray}{rCl}
\lim_{K\to\infty} \frac{|\Tf|}{K} & = & \frac{| \mathcal{T}_{\text{subnet}} \cap \Tf| }{s}  = \frac{\D^2 - 2 \D +4}{3 \D^2}\\
\lim_{K\to\infty} \frac{|\Ts|}{K} & = &  \frac{| \mathcal{T}_{\text{subnet}} \cap \Ts| }{s}= \frac{2 \D - 4 }{3 \D}.
\end{IEEEeqnarray}
By \eqref{eq:MG}, this establishes  the achievability of  the desired MG pair in \eqref{both-hexa}. 
\begin{lemma}\label{eq:lemma_sub}
The number of ``fast" Txs in  $\mathcal{T}_{\textnormal{subnet}}$ equals 
\begin{equation}\label{tfnum}
| \mathcal T_{\text{fast}} \cap \mathcal T_{\textnormal{subnet}}| = \frac{\D^2}{4}-\frac{\D}{2} + 1, 
\end{equation}
and the number of ``slow" Txs in $\mathcal T_{\textnormal{subnet}}$
equals
\begin{equation} \label{tsnum}
  | \mathcal T_{\text{slow}} \cap \mathcal T_{\textnormal{subnet}}| = \frac{\D^2}{2} - \D.
\end{equation}
\end{lemma}
\begin{IEEEproof}
For ease of notation, assume that $k_{\textnormal{master}}=0$.    

Define the sector of the subnet $\mathcal{T}_{\textnormal{subnet}}$ with positive coordinates $a_k>0$ and $b_k \geq 0$ (recall that we assume $k_{\textnormal{master}}=0$):
\begin{equation} \label{eq:tplus}
\mathcal T_{\text{subnet}}^+ \triangleq \left \{k \colon  0<a\le \frac{\D}{2} -1, \; 0\le b \le  \frac{\D}{2}-1\right \} \subseteq \mathcal{T}_{\text{subnet}}.
\end{equation}
Since the angle between the coordinate vectors $\mathbf{e}_x$ and $\mathbf{e}_y$ is $\frac{2 \pi}{3}$ and the sets $\mathcal{T}_{\textnormal{fast}}$ and $\mathcal{T}_{\textnormal{subnet}}$ are rotationally-invariant with respect to this angle, 
\begin{equation}\label{tfnum}
| \mathcal T_{\text{subnet}} \cap \mathcal T_{\textnormal{fast}}| = 3 | \mathcal T_{\text{subnet}}^+ \cap \mathcal T_{\textnormal{fast}}|+1, 
\end{equation}
where the $1$  has to be added to account for the ``fast" master cell at the origin.

To calculate the  size of the set $\mathcal T_{\text{subnet}}^+ \cap \mathcal T_{\textnormal{fast}}$, notice that by Assumption \eqref{eq:59},  $\frac{\D}{2}-1$ is a multiple of $3$ and thus for each value of $b\in\{0,1,\ldots, \frac{D}{2}-1\}$ there are exactly $\frac{\frac{\D}{2}-1}{3}$ values $a\in \{1,\ldots, \frac{\D}{2}-1\}$ so that the sum $a+b$ is a multiple of $3$. 
Therefore, 
\begin{equation}
|\mathcal T_{\text{fast}} \cap \mathcal{T}_{\text{subnet}}^+| = \frac{\D}{2} \cdot \frac{\frac{\D}{2}-1}{3}, 
\end{equation}
and 
\begin{equation}\label{eq:fast_subnet}
| \mathcal T_{\text{fast}} \cap \mathcal T_{\text{subnet}}| = 3   |\mathcal T_{\text{fast}} \cap \mathcal{T}_{\text{subnet}}^+|  + 1 = \frac{\D^2}{4}-\frac{\D}{2} + 1.
\end{equation}

Since all cells in $\mathcal{T}_{\text{subnet}}$ that are not elements of $\mathcal{T}_{\text{fast}}$ belong to $\mathcal{T}_{\text{slow}}$, 
we obtain by \eqref{eq:size_subnet} and \eqref{eq:fast_subnet}  that 
\begin{equation}
|\Ts \cap \mathcal{T}_{\text{subnet}}|=| \mathcal{T}_{\text{subnet}}| - |\Tf \cap\mathcal{T}_{\text{subnet}}|= \left(1 + \frac{3}{4} \D (\D-2) \right)- \left(  \frac{\D^2}{4}-\frac{\D}{2} + 1\right) = \frac{\D^2}{2} -\D
\end{equation}
\end{IEEEproof}

We  analyze the required cooperation prelogs. As in the previous subsection, $\gamma_{\Tx,k}=\gamma_{\Rx,k}=i$ for every  cell~$k$ that lies $i$ cell hops away from its next master cell, for $i=1, \ldots, \D/2-1$. Moreover, for each ``fast'' Tx~$k$ with $\gamma_{\Tx,k} \in \{1, \ldots, \D/2 -2\}$ the size of the ``slow'' interfering set $\mathcal I_k^{(S)}$ is equal to $6$, and when $ \gamma_{\Tx,k} = \D/2-1$ the size of this set is equal to $3$ for the corner ``fast''-cells and it is equal to $4$ for the other ``fast''-cells of this layer.  
By Assumption~\eqref{eq:59}, the 6 corner  cells  are all ``fast" cells. In fact, they are given by $(a_k=\frac{\D}{2}-1, b_k=0)$, $(a_k=\frac{\D}{2}-1, b_k=\frac{\D}{2}-1)$, $(a_k=0, b_k=\frac{\D}{2}-1)$, $(a_k=-\frac{\D}{2}+1, b_k=0)$, $(a_k=-\frac{\D}{2}+1, b_k=-\frac{\D}{2}+1)$, $(a_k=0, b_k=-\frac{\D}{2}+1)$, for which $a_k+b_k$ is a multiple of $\frac{\D}{2}-1$ and by  \eqref{eq:59}   divisible by $3$. Between any two corner cells there are $\frac{\D}{2} -2$ cells with $\gamma_{\Tx,k}=\frac{\D}{2}-1$, and  $(\frac{\D}{2}-4)/3$ of them are fast cells. (This can be seen by the previously mentioned rotation invariance of both sets $\mathcal{T}_{\text{subnet}}$ and $\Tf$ with respect to the angle $\frac{2\pi}{3}$, and because the non-corner cells in $\mathcal{T}_{\textnormal{subnet}}^+$ with $\gamma_{\Tx,k}=\frac{\D}{2}-1$    either  have coordinates $a_k=\frac{\D}{2}-1$ and $b_k=1, \ldots, \frac{\D}{2}-2$ or they have coordinates $a_k=1, \ldots, \frac{\D}{2}-2$ and $b_k=\frac{\D}{2}-1$. It is easily seen that since $\frac{\D}{2}-1$ is a multiple of $3$, a total number of $2\big(\frac{\D}{2}-4\big)/3$ of these cells have sum $a_k+b_k$ that is divisible by $3$.)

Since there are $6$  ``fast'' corner cells with $\gamma_{\Tx,k}=\D/2-1$,  we conclude that there are $6 \cdot \big(\frac{\D}{2}-4\big)/3$ ``fast" non-corner cells with $\gamma_{\Tx,k}=\D/2-1$. 
Combining these considerations with \eqref{eq:fast_subnet}, 
we  conclude  that 
\begin{IEEEeqnarray}{rCl}
\lefteqn{\sum_{k \in \mathcal{T}_{\text{fast}} \cap \mathcal{T}_{\textnormal{subnet}}}\left  |\mathcal{I}_{k}^{(S)}\right | }  \qquad \nonumber \\
&=&  \sum_{\substack{k \in \mathcal{T}_{\text{fast}} \cap \mathcal{T}_{\textnormal{subnet}} \colon \\ \gamma_{\Tx, k } \neq \frac{\D}{2} -1} }6  + \sum_{\substack{k \in \mathcal{T}_{\text{fast}} \cap \mathcal{T}_{\textnormal{subnet}}\colon  \\ \gamma_{\Tx, k } =\frac{\D}{2} -1 \\ k \text{ a non-corner cell}} }4 + \sum_{\substack{k \in \mathcal{T}_{\text{fast}} \cap \mathcal{T}_{\textnormal{subnet}} \colon  \\ \gamma_{\Tx, k } =\frac{\D}{2} -1  \\ k \text{ a corner cell} }}3 \\
& = & 6\cdot  | \mathcal{T}_{\text{fast}} \cap \mathcal{T}_{\textnormal{subnet}}| - 2\cdot \left |\left \{k \in  \mathcal{T}_{\text{fast}} \cap \mathcal{T}_{\textnormal{subnet}}\colon  \gamma_{\Tx, k } =\frac{\D}{2} -1 \text{ and } k \text{ a non-corner cell} \right \} \right |  \nonumber \\
& & - 3 \cdot \left  | \left \{k \in  \mathcal{T}_{\text{fast}} \cap \mathcal{T}_{\textnormal{subnet}}\colon  \gamma_{\Tx, k } =\frac{\D}{2} -1 \text{ and } k \text{ a corner cell} \right\}   \right |  \\
& = & 6 \cdot \left(\frac{\D^2}{4}-\frac{\D}{2} + 1\right) - 2 \cdot 6 \cdot \frac{\frac{\D}{2}-4}{3}- 3 \cdot 6 \label{eq:inter}\\
&=& \frac{(3 \D-4)(\D-2)}{2} = \frac{3 \D^2}{2} - 5 \D +4 .\label{eq:sum_I}
\end{IEEEeqnarray}
 Since $\frac{\q_{K, \Rx}}{K}  \to 6$ and since, by a sandwiching argument, $\frac{1}{K} \sum_{k \in \mathcal{T}_{\text{fast}} } |\mathcal{I}_{k}^{(S)} |  \to s^{-1}\sum_{k \in \mathcal{T}_{\text{fast}} \cap \textnormal{subnet}}  |\mathcal{I}_{k}^{(S)} |$ as $K\to \infty$: 
 \begin{IEEEeqnarray}{rCl}
 \lim_{K\to \infty} \frac{\sum_{k \in \mathcal{T}_{\text{fast}} }\left  |\mathcal{I}_{k}^{(S)}\right | }{ \q_{K,\Rx}}=  \lim_{K\to \infty} \frac{\sum_{k \in \mathcal{T}_{\text{fast}} \cap \mathcal{T}_{\textnormal{subnet}}}\left  |\mathcal{I}_{k}^{(S)}\right | }{s \q_{K,\Rx}/K} = \frac{\frac{(3 \D-4)(\D-2)}{2}}{ \frac{3}{4}\D^2\cdot 6}= \frac{ (3 \D-4)(\D-2)}{9 \D^2}.
\end{IEEEeqnarray}
 Then, by \eqref{mtbone}:
\begin{equation}\label{eq:res_mt}
\mtboneh  =   \L \cdot \frac{(3 \D-4)(\D-2)}{9\D^2}.
\end{equation}

 To calculate the Rx-cooperation prelog, we notice that for each  ``slow'' cell~$k$ with $\gamma_{\Rx,k} \in \{1, \ldots, \D/2 -2\}$  the ``fast'' interfering set $\mathcal I_{k}^{(F)}$  is of size $3$, and for each  ``slow'' cell~$k$ with $\gamma_{\Rx,k} = \D/2-1$ it is of size $2$. As explained previously, among the $6\cdot \big(\frac{\D}{2}-1\big)$ cells with $\gamma_{\Rx,k}=\frac{\D}{2}-1$, there are $6$ ``fast" corner cells and $\D-8$ ``fast" non-corner cells.   The remaining are ``slow" cells, and therefore 
\begin{IEEEeqnarray}{rCl}
 \sum_{k \in \mathcal{T}_{\text{slow}}\cap \mathcal{T}_{\textnormal{subnet}}}  |\mathcal{I}_{k}^{(F)}|  
 & = & 3 | \mathcal{T}_{\text{slow}}\cap \mathcal{T}_{\textnormal{subnet}}| - | \{ k\in (\mathcal{T}_{\text{slow}}\cap \mathcal{T}_{\textnormal{subnet}}) \colon \gamma_{\Tx,k}=\D/2-1\} | 
\\
&=& 3 \left( \frac{\D^2}{2} -\D\right) -\left(  6 \cdot \left(\frac{\D}{2}-1\right) - 6 - ( \D -8)\right)= \frac{3 \D^2}{2} - 5 \D +4. \label{eq:sumIkf} \IEEEeqnarraynumspace
\end{IEEEeqnarray}
 
To calculate the sum $\sum_{k \in \mathcal{T}_{\text{slow}}\cap \mathcal{T}_{\textnormal{subnet}}} \gamma_{\Rx,k}$, we first characterize the number of ``fast" cells in  the sector $\mathcal{T}_{\text{subnet}}^+$ that have $\gamma_{\Rx,k}=i$, for $i=1,\ldots \D/2-1$. A cell $k\in \mathcal{T}_{\textnormal{subnet}}^+$ has $\gamma_{\Rx,k}=i$  if it has coordinates of the form  $a_k=i$ and $b_k = 0, \ldots, i-1$ or    $a_k=1,\ldots, i$ and $b_k=i$.  To count the number  of ``fast" cells in this set, we distinguish different cases for $i\mod 3$:
\begin{itemize}
\item If $i\mod 3=0$, then there are $2 i/3$ ``fast" cells among these cells (namely the cells $(a_k=i, b_k=0)$, $(a_k=i, b_k=3), \ldots, (a_k=i, b_k=i)$, $(a_k=i-3, b_k=i), \ldots, (a_k=3, b_k=i)$) and the remaining $4i/3$ cells are ``slow" cells. 

\item If $i \mod 3=1$, then  there are $2 (i-1)/3$ ``fast" cells among these cells (namely the cells $(a_k=i, b_k=2)$, $(a_k=i, b_k=5), \ldots, (a_k=i, b_k=i-2), (a_k=i-2, b_k=i), \ldots, (a_k=2, b_k=i)$) and the remaining $2(2i+1)/3$ cells are ``slow" cells.
 
\item If $i \mod 3=2$, then  there are $2 (i+1)/3$ ``fast" cells among these cells (namely the cells $(a_k=i, b_k=1)$, $(a_k=i, b_k=4), \ldots, (a_k=i, b_k=i-1), (a_k=i-1, b_k=i), \ldots, (a_k=1, b_k=i)$) and the remaining $2(2i-1)/3$ cells are ``slow" cells. 
\end{itemize}
By the rotation invariance of all relevant sets with respect to the angle $2\pi/3$, we  conclude that 
 \begin{equation}\label{tsii}
\left |\{ k\in \Ts \cap \mathcal T_{\textnormal{subnet}}: \gamma_{\Rx,k} = i\}\right | =\begin{cases}
 4i& \quad i \mod 3 = 0,\\
 4i+2&  \quad i \mod 3 = 1,\\
 4i-2& \quad i \mod 3 = 2.
 \end{cases} 
\end{equation}
and since $\frac{\D}{2}-1$ is a multiple of $3$, see \eqref{eq:59}:
 \begin{IEEEeqnarray}{rCl}
\sum_{k \in \mathcal{T}_{\text{slow}}\cap \mathcal{T}_{\textnormal{subnet}}} \gamma_{\Rx,k} 
& = & \sum_{i=1}^{\frac{\D}{2}-1} 4 i^2 + 2  \sum_{j=1}^{\frac{\frac{\D}{2}-1}{3}} ( 3 j -2) -2 \sum_{j=1}^{\frac{\frac{\D}{2}-1}{3}} ( 3 j -1) \\
& = &  \frac{\D^3 - 3\D^2 + 4}{6}.\label{eq:sum_g}
	\end{IEEEeqnarray}

 Putting \eqref{eq:sum_I} and \eqref{eq:sum_g} together, we obtain: 
\begin{IEEEeqnarray}{rCl}
\sum_{k \in \mathcal{T}_{\text{slow}}\cap \mathcal{T}_{\textnormal{subnet}}} \left( |\mathcal{I}_{k}^{(F)}| + 2\gamma_{\Rx,k} \right)
	&=& \frac{3 \D^2}{2} - 5 \D +4  + 2 \cdot \left(\frac{\D^3 - 3\D^2 + 4}{6} \right)  \\ 
	&=& \frac{2\D^3 + 3\D^2 - 30\D +32}{6}. \label{eq:d}
\end{IEEEeqnarray}
 Since $\frac{\q_{K, \Rx}}{K}  \to 6$ and, by a sandwiching argument,  $K^{-1}\sum_{k \in \mathcal{T}_{\text{slow}} } |\mathcal{I}_{k}^{(F)} |  \to s^{-1}\sum_{k \in \mathcal{T}_{\text{slow}} \cap \mathcal{T}_{\text{subnet}}}  |\mathcal{I}_{k}^{(F)} |$ as $K\to \infty$: 
 \begin{IEEEeqnarray}{rCl}
 \lim_{K\to \infty} \frac{\sum_{k \in \mathcal{T}_{\text{slow}} }\left  |\mathcal{I}_{k}^{(F)}\right | + 2\gamma_{\Rx,k} }{ \q_{K,\Rx}} & =&   \lim_{K\to \infty} \frac{\sum_{k \in \mathcal{T}_{\text{fast}} \cap \mathcal{T}_{\textnormal{subnet}}}\left  |\mathcal{I}_{k}^{(S)}\right | + 2\gamma_{\Rx,k} }{s \q_{K,\Rx}/K} \nonumber \\
 & = & 
 \frac{2\D^3 + 3\D^2 - 30 \D +32}{6 \cdot \frac{3}{4} \D^2 \cdot 6}\\
 &=&  \frac{2\D^3 + 3\D^2 - 30 \D +32}{27 \D^2}.
 \end{IEEEeqnarray}

 By \eqref{mrbone}, the average Rx-cooperation prelog  is 
\begin{equation}
\mrboneh  =   \L \cdot \frac{2\D^3 + 3\D^2 - 30 \D +32}{27\D^2}.
\end{equation}

\subsection{Coding scheme to transmit both ``fast'' and ``slow'' messages with CoMP transmission} \label{sub:hexa-bothT}
We choose the same  cell association as in the previous subsection. Consequently,  the scheme achieves the same MG pair and 
 the single-round Rx-cooperation prelog coincides with the single-round Tx-cooperation prelog  in the previous  Subsection \ref{sub:hexa-bothR} (see \eqref{eq:res_mt}):
\begin{equation}
\mrbtwoh = \mtboneh.
\end{equation} 
To calculate the average Tx-cooperation prelog, we first consider  the $q$-term in \eqref{mtbtwo},  which characterizes the number of quantization messages describing the ``slow" signals that are counted twice: once for the CoMP transmission and once for the interference mitigation at ``fast" Txs. Since the master Tx is a ``fast''-Tx, all  $6$ incoming messages are counted twice. Moreover, for each ``fast''-Tx~$k$ in layer $i$ (i.e., with $\gamma_{\Tx,k}=i$), for $i \in \{1,\ldots, \frac{\D}{2} -2\}$,  there are $2$ neighbouring ``slow''  Txs in the subsequent layer $i +1$. Thus for each such ``fast'' Tx, there are $2$ messages that are double-counted. 
Repeating the  arguments that  justify \eqref{eq:inter}, we obtain that 
\begin{equation} 
| \{ k\in \mathcal{T}_{\text{subnet}}\cap \Tf \colon \gamma_{\Tx,k}=i \}| = \left( \frac{\D^2}{4}- \D \right) -  (6+ \D-8 )= \frac{\D^2}{4} - \frac{3 \D }{2} +2.
\end{equation}

Therefore, 
\begin{equation}
q_{\text{subnet}} = 6 + 2 \left( \frac{\D^2 - 6\D + 12}{4} -1\right) = \frac{\D^2}{2} - 3\D + 10.
\end{equation}
The sum $\sum_{k \in \mathcal T_{\text{slow}} \cap \mathcal T_{\textnormal{subnet}}} 2 \gamma_{\Tx, k} + \sum_{k \in \mathcal T_{\text{fast}} \cap \mathcal T_{\textnormal{subnet}}} |\mathcal I_k^{(S)}|$ can be calculated as in the previous subsection for the scheme with CoMP reception. Specifically,  since  $\gamma_{\Tx,k} = \gamma_{\Rx,k}$,  the sum $\sum_{k \in \mathcal T_{\text{slow}} \cap \mathcal T_{\textnormal{subnet}}} \gamma_{\Tx, k}$ is given by  the right-hand side  of \eqref{eq:d}.  The sum $ \sum_{k \in \mathcal T_{\text{fast}} \cap \mathcal T_{\textnormal{subnet}}} |\mathcal I_k^{(S)}|$ is calculated in \eqref{eq:sum_I}. This establishes:
\begin{equation} \label{eq:sum-q}
\sum_{k \in \mathcal T_{\text{slow}} \cap \mathcal T_{\textnormal{subnet}}} 2 \gamma_{\Tx, k} + \sum_{k \in \mathcal T_{\text{fast}} \cap \mathcal T_{\textnormal{subnet}}} |\mathcal I_k^{(S)}| -q_{\textnormal{subnet}} = \frac{2\D^3 -12\D -28}{6}.
\end{equation}

By now  standard asymptotic  arguments and by \eqref{mtbtwo}, the average Tx-cooperation prelog is then obtained by dividing \eqref{eq:sum-q} by $s=\frac{3}{4}\D^2$ and multiplying it by the number of antennas $\L$:
\begin{equation}
\mtbtwoh  =   \L \cdot \frac{2\D^3 -12\D -28}{27\D^2}.
\end{equation}

\section{Coding Schemes and Analysis in the Sectorized Hexagonal Model} \label{app-B}
  In this appendix we prove that the Tx/Rx set associations  proposed in  Subsection~\ref{sub:sectset} are permissible and we provide details on how to compute the  corresponding MG pairs and cooperation prelogs.
   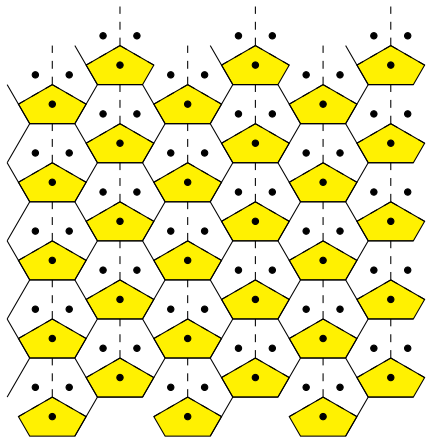
\begin{figure}[t]
  \vspace*{-2ex}
  	\centering
  	\footnotesize
  	\begin{tikzpicture}[scale=0.6, >=stealth]
  	\centering
  	{
  	\foreach \m in {-1,0,1,2,3}
  	\foreach \i in {0,1,2}{
  	 \draw [fill = yellow](0.5+3*\i+1.5,0.8660+1.7321*\m+0.8660 + 1.7321)-- +(-30:0.8660)--(3*\i+1+1.5,2*sin{60}*3+0.8660+ 1.7321*\m - 2*1.7321)--(3*\i+1.5,2*sin{60}*3+0.8660+ 1.7321*\m - 2*1.7321) --+(120:0.5)--(0.5+3*\i+1.5,0.8660+1.7321*\m+0.8660+ 1.7321);}
  	\foreach \m in {-1,0,1,2,3}
  	\foreach \i in {0,1,2}{ 
  	  \draw [fill = yellow] (0.5+3*\i,0.8660+1.7321*\m+ 1.7321)-- +(-30:0.8660)--(3*\i+1,2*sin{60}*3+ 1.7321*\m - 2*1.7321)--(3*\i,2*sin{60}*3+ 1.7321*\m-2*1.7321) --+(120:0.5)--(0.5+3*\i,0.8660+1.7321*\m+ 1.7321);}
  	  }
  	 \foreach \i in {0,...,2} 
  \foreach \j in {1,...,4} {
  \foreach \a in {0,120,-120} \draw[black] (3*\i,2*sin{60}*\j) -- +(\a:1);}
   \foreach \i in {0,...,2} 
  \foreach \j in {0,...,4} {
 \foreach \a in {0,120,-120} \draw (3*\i+3*cos{60},2*sin{60}*\j+sin{60}) -- +(\a:1);}
  \foreach \ii in {0,...,4}
    \foreach \jj in {0,...,2}{
  \foreach \a in {90,-30,-150} \draw[black,dashed] (0.5+3*\jj,0.8660+\ii*1.7321) -- +(\a:0.8660);
    \foreach \a in {90,-30,-150} \draw[black,dashed] (2+3*\jj,1.7321+\ii*1.7321) -- +(\a:0.8660);}
  \foreach \c in {0,...,2}
     \foreach \cc in {0,...,4}{
 \draw [fill=black] (0.5 + 3*\c, 1.7321/4+1.7321*\cc) circle (0.07);}
  \foreach \c in {0,...,2}
     \foreach \cc in {0,...,4}{
 \draw [fill=black] (2+ 3*\c, 3*1.7321/4+1.7321*\cc) circle (0.07);}
 \foreach \ccc in {0,...,2}
     \foreach \cccc in {0,...,4}{
  \draw [fill=black] (0.5+0.5*0.75+ 3*\ccc, 1.7321/2+1.7321/8+1.7321*\cccc) circle (0.07);
   \draw [fill=black] (2-0.5*0.75+3*\ccc, 1.7321/2+1.7321/2+1.7321/8+1.7321*\cccc) circle (0.07);
     \draw [fill=black] (0.5-0.5*0.75+ 3*\ccc, 1.7321/2+1.7321/8+1.7321*\cccc) circle (0.07);
   \draw [fill=black] (2+0.5*0.75+3*\ccc, 1.7321/2+1.7321/2+1.7321/8+1.7321*\cccc) circle (0.07);}

  	\end{tikzpicture} 
  	\caption{Illustration of the scheme without cooperation in the sectorized hexagonal network.}
  	\label{fig14-2}
  \end{figure}~~%

  \subsection{No cooperation scheme}
  Figure~\ref{fig14-2} shows the active Txs in yellow and the silenced Txs in white.  It is easily seen that transmissions in yellow sectors do not interfere, as they pertain to non-neighbouring sectors. This scheme requires no cooperation messages and  achieves the sum MG in \eqref{sec-fast}. 
  
  \subsection{Coding scheme to transmit only ``slow'' messages with CoMP reception} \label{sub:sec-slow}
  
\maw{The proposed cell association splits the network into non-interfering subnets, because all sectors in layer $\D/2$ that belong to two different subnets are silenced.}
Moreover, in layer $\D/2$ \maw{of a given subnet,  3 of the 6 corner cells} are completely silenced and each of the remaining cells contains exactly one active sector pertaining to the subnet. The total number of active sectors per cell is thus
\begin{equation}\label{eq:s_active}
s_{\textnormal{active}}=3+  3\sum_{i= 1}^{\D/2-1} 6 i  + (3\D-3) = 9\D^2/4 - 3\D/2 .
\end{equation}
A valid sector partitioning can be obtained  by associating all sectors of the cell partitioning proposed for  the hexagonal model to the same subset, which then contains (by \eqref{eq:subnet_size2} and because each cell contains 3 sectors):
\begin{equation}\label{eq:s_sectors}
s_{\textnormal{sectors}} = 3 \frac{3}{4} \D^2.
\end{equation}
%

By these considerations
\begin{equation}\label{lim:slow22}
\lim_{K\to \infty} \frac{|\mathcal T_{\text{slow}}|}{\maw{3K}}= \frac{s_{\textnormal{active}}}{s_{\textnormal{sectors}}} =\frac{9\D^2/4 - 3\D/2 }{9\D^2/4}= \frac{3\D-2}{3\D},
\end{equation}
and as a result, by \eqref{mgs} and \eqref{ssm}, the proposed cell association achieves the MG pair $(\S^{(F)}=0, \S^{(S)}=\S^{(S)}_{\textnormal{max}})$ with $\S^{(S)}_{\textnormal{max}}$ defined in \eqref{slow-sec}.

With CoMP reception, this scheme does not use any Tx-cooperation messages and $\mtstwoh=0$. To calculate the required Rx-cooperation prelog, notice that   for each \maw{created  \emph{sector}-subnet $\mathcal{T}_{\text{subnet}}\in[3K]$ and each sector~$k \in \mathcal{T}_{\textnormal{subnet}}$, $\gamma_{\Rx,k}=i$ if $k$ lies in the $i$-th \emph{cell-layer}} around the master cell. Since in each layer $i\in\{1,\ldots, \D/2-1\}$ there are $6i$ cells and thus $18i$ sectors and in layer $\D/2$ there are $3\D-3$ active sectors as explained above, 
\begin{equation}
2\sum_{k \in  \mathcal{T}_{\text{subnet}}} \gamma_{\Rx,k} = 2\cdot \left(\sum_{i =1}^{\D/2-1} (18 i)\cdot i+  (3\D-3)\cdot \D/2 \right)= \frac{3\D^2(\D-1)}{2},
\end{equation}
and by a sandwiching argument
\begin{equation}
\lim_{K\to \infty}  \frac{2\sum_{k \in  \mathcal{T}_{\textnormal{slow}}} \gamma_{\Rx,k} }{\maw{3K}} =  \frac{2\sum_{k \in  \mathcal{T}_{\text{subnet}}} \gamma_{\Rx,k} }{s_{\textnormal{sector}}}=\frac{ (3\D^2(\D-1))/2}{\maw{\frac{9\D^2}{4}}} = \maw{\frac{2(\D-1)}{3}}.
\end{equation}
Since \maw{$\lim_{K\to \infty} \frac{\q_{K, \Rx}}{{3K}} =2$},  according to \eqref{mrstwo}
 the required Rx-cooperation prelog  equals   
\begin{equation}
\mrstwoh = \L \cdot  \frac{(\D-1)}{3}.
\end{equation}

%
  
\subsection{Coding scheme to transmit both ``fast'' and ``slow'' messages with CoMP reception}  \label{sub:sec-bothR}
Consider  the cell and sector association for this network described in the main body  of the paper and illustrated in Fig.~\ref{fig5.10b} for $\D = 8$ where  white sectors are deactivated, Txs in yellow sectors send ``fast'' messages, and Txs in blue sectors send ``slow'' messages. As explained in the preceeding subsection,  the network is split into subnets and the total number of active sectors in a subnet  is $9\D^2/4 - 3\D/2$, see \eqref{eq:s_active}. 

We count the number of ``fast" sectors  in a subnet $\mathcal T_{\text{subnet}}$ surrounding a master cell at the origin. Notice that all subnets are symmetric and have same number of ``fast" and ``slow" sectors. 
As explained in the main body of the paper, each cell in layer $\D/2$ has 1 active ``fast" sector pertaining to the subnet, except for 3 three corner cells that are completely desactivated. The number of ``fast" sectors in layer $\D/2$ is thus $3 \D -3$. In each layer-$i$ with $i \in \{1, \ldots, \frac{\D}{2}-1\}$, each cell has exactly one ``fast' sector, except the cells with coordinates satisfying one of the three conditions: ($a_k=0$ and $b_k > 0$) or ($a_k> 0$ and $b_k=0$) or ($a_k=b_k < 0$). There are $3(\D/2-1)$ such cells, and thus the total number of ``fast" sectors in layers $i=1,\ldots, \D/2-1$ is:
\begin{equation}
\sum_{i=1}^{\D/2-1} 6i -3(\D/2-1) =3  \frac{\D(\D-2)}{4} -\frac{3\D}{2} +3 = \frac{3}{4} \D^2 -3 \D +3.
\end{equation}  Since the master cell sends ``slow" messages only, we obtain that the number of ``fast'' sectors in subnet $\mathcal T_{\text{subnet}}$ equals
\begin{equation}\label{eq:134}
\left | \Tf \cap \mathcal T_{\text{subnet}} \right | = (3 \D-3) + \left( \frac{3}{4} \D^2 -3 \D +3 \right) =  \frac{3}{4} \D^2 .
\end{equation}
Since the total number of active Txs in this subnet equals $s_{\textnormal{active}}=\frac{9\D^2}{4}-\frac{3\D}{2}$, see \eqref{eq:s_active}, the subnet's number of ``slow'' Txs is:
\begin{equation}
\left | \Ts \cap \mathcal T_{\text{subnet}} \right | = \frac{6\D^2}{4} - \frac{3\D}{2}.
\end{equation}
We notice that similarly to the previous subsection, one can obtain a valid cell partitioning by associating a subset of $s= 3 \frac{3}{4} \D^2$ cells to each master cell.  Since each cell has $3$ sectors and because all subnets are equal,  applying standard sandwiching arguments to eliminitate edge effects for finite number of users $K$, we  obtain:
\begin{equation}
\lim_{K\to \infty} \frac{ | \mathcal T_{\text{fast}} |}{\maw{3}K}  =  \frac{\left | \Tf \cap \mathcal T_{\text{subnet}} \right | }{3s}= \frac{\frac{3\D^2}{4}}{\frac{9 \D^2}{4}} =  \frac{1}{3} 
\end{equation}
and 
\begin{equation}
\lim_{K\to \infty} \frac{ | \mathcal T_{\text{slow}} |}{\maw{3}K}  = \frac{ \left | \Ts \cap \mathcal T_{\text{subnet}} \right |  }{3 s}= \frac{\frac{6\D^2}{4} - \frac{3\D}{2}}{\frac{9 \D^2}{4}} =  \frac{2\D - 2}{3 \D}.
\end{equation}
\maw{T}his establishes  the achievability of  the MG pair
\eqref{both-sect}.

%
%
%
%
%
%

 To analyze the cooperation prelog of the sector association,  notice  that for each ``fast'' Tx/Rx~$k$ with $\gamma_{\Rx,k} \in \{1, \ldots, \D/2 -1\}$ the size of the ``slow'' interfering set $\mathcal I_k^{(S)}$ is equal to $4$, and when $ \gamma_{\Rx,k} =\D/2$ the size of this set is equal to $2$ for the three active corner cells and equal to $3$ for the other non-corner cells. Considering the fact that the number of ``fast'' Txs with $\gamma_{\Rx,k} = \frac{\D}{2}$ equals $3\D-3$,  then by \eqref{eq:134} we conclude that 
 
  \begin{equation}
\sum_{k \in \mathcal{T}_{\text{fast}} \cap \mathcal{T}_{\text{subnet}}} |\mathcal{I}_{k}^{(S)}| = 4\left ( \frac{3\D^2}{4}-(3\D-3)\right ) + 2 \cdot 3 + 3(3\D-6) = 3\D(\D-1).
\end{equation}

   According to \eqref{mtbone} \maw{and sandwiching arguments,} the average Tx-cooperation prelog required for the scheme is
\begin{equation}
  \maw{\mtbones  =  \L  \cdot  \frac{\frac{\sum_{k \in \mathcal{T}_{\text{fast}} \cap \mathcal{T}_{\text{subnet}}} |\mathcal{I}_{k}^{(S)}|}{3s} }{\lim_{K\to \infty} \frac{\q_{K, \Tx}}{{K}}} =\L \cdot  \frac{ \frac{ 3\D(\D-1)}{9/4 \cdot \D^2}}{4}=  \L \cdot \frac{(\D-1)}{3\D}.}
  \end{equation}

 Fig.~\ref{fig5.10b} also shows that  the size of the ``fast'' interference set $\mathcal I_{k}^{(F)}$  is equal to $2$ for each  ``slow'' Tx~$k$. To precisely calculate the number of required Rx-cooperation messages, notice that each non-corner ``fast''-Rx~$k$ with $\gamma_{\Rx,k} = \frac{\D}{2}$ sends its decoded message to two of its neighbours and each corner ``fast''-Rx~$k$ with $\gamma_{\Rx,k} = \frac{\D}{2}$ sends its decoded message to only one neighbouring Rx. As there are $3\D-6$ non-corner Rxs and $3$ active corner Rxs in this layer,  the total number of cooperation messages by these Rxs equals $6\D - 9$. Any other ``fast'' Rx \maw{that} is not in this layer has to send its decoded messages to $3$ of its neighbours. By \eqref{eq:134},  these ``fast''  Rxs send in total  $3(\frac{3\D^2}{4} - (3\D-3))$ cooperation messages to their neighbours. Each Rx decoding a ``slow'' message also sends the quantized version of its channel outputs to the next master Rx.  Among the Rxs with $\gamma_{\Rx,k} = i$, there are $6i-3$ Rxs observing two ``slow'' signals and $3$ Rxs observing $3$ ``slow'' signals. To sum up, the total number of Rx-cooperation messages transmitted \maw{within a subnet} is 
  \begin{IEEEeqnarray}{rCl}
\sum_{k \in \mathcal{T}_{\text{slow}}\cap \mathcal{T}_{\text{subnet}}} |\mathcal{I}_{k}^{(F)}| +2\gamma_{\Rx,k} &=& 6\D-9+ 3\left(\frac{3\D^2}{4} - (3\D-3)\right) + 4\sum_{i=1}^{\frac{\D}{2}-1}i(6i-3) + 6\sum_{i=1}^{\frac{\D}{2}-1}3i \notag \\
&=& \frac{\D(2\D^2-5)}{2}.  
\end{IEEEeqnarray}
  Thus according to \eqref{mrbone}  \maw{and sandwiching arguments,} the average Rx-cooperation prelog required by the scheme is 
\begin{equation}
\maw{  \mrbones  =    
  \L  \cdot  \frac{\frac{\sum_{k \in \mathcal{T}_{\text{slow}}\cap \mathcal{T}_{\text{subnet}}} |\mathcal{I}_{k}^{(F)}| +2\gamma_{\Rx,k}}{3 s} }{\lim_{K\to \infty} \frac{\q_{K, \Rx}}{{3K}}} =\L \cdot  \frac{  \frac{{\D(2\D^2-5)}/{2}}{9/4 \cdot \D^2}}{2}=
  \L \cdot \frac{2\D^2-5}{9\D}.  }
  \end{equation}
 \end{document}